\documentclass[preprint,aps,prd,showpacs,nofootinbib]{revtex4}%
\usepackage{graphicx}
\usepackage{amsmath}
\usepackage{amssymb}
\usepackage{bm}
\usepackage{epsfig}
\def\sl{\!\!\!/}
\begin{document}
\preprint{}
\title{\mbox{}\\[10pt]
Double logarithms in $\bm{e^+ e^- \to J/\psi + \eta_c}$
}
\author{Geoffrey~T.~Bodwin}
\affiliation{High Energy Physics Division, Argonne National Laboratory,\\
9700 South Cass Avenue, Argonne, Illinois 60439, USA}
\author{Hee~Sok~Chung}
\affiliation{High Energy Physics Division, Argonne National Laboratory,\\
9700 South Cass Avenue, Argonne, Illinois 60439, USA}
\author{Jungil~Lee}
\affiliation{Department of Physics, Korea University, Seoul 136-713, Korea}
\date{\today}
\begin{abstract}
Double logarithms of $Q^2/m_c^2$ that appear in the cross section for
$e^+e^-\to J/\psi + \eta_c$ at next-to-leading order (NLO) in the strong
coupling $\alpha_s$ account for the bulk of the NLO correction at
$B$-factory energies. (Here, $Q^2$ is the square of the
center-of-momentum energy, and $m_c$ is the charm-quark mass.) We
analyze the double logarithms that appear in the contribution of each 
NLO Feynman diagram, and we find that the double logarithms arise from
both the Sudakov and the end-point regions of the loop integration.
The Sudakov double logarithms cancel in the sum over all diagrams.  We
show that the end-point region of integration can be interpreted as
a pinch-singular region in which a spectator fermion line becomes soft or
soft and collinear to a produced meson. This interpretation may be
important in establishing factorization theorems for helicity-flip
processes, such as $e^+e^-\to J/\psi + \eta_c$, and in resumming
logarithms of $Q^2/m_c^2$ to all orders in $\alpha_s$.
\end{abstract}
\maketitle
\section{Introduction\label{sec:intro}}

Measurements of the exclusive double-charmonium production cross section
$\sigma (e^+ e^- \to J/\psi + \eta_c)$ by the Belle \cite{Abe:2002rb,
Abe:2004ww} and {\it BABAR} \cite{Aubert:2005tj} collaborations have
stimulated a good deal of theoretical activity. Calculations within the
nonrelativistic QCD (NRQCD) factorization formalism \cite{Bodwin:1994jh}
at leading order (LO) in $\alpha_s$ and $v$ \cite{Braaten:2002fi,
Liu:2002wq} yield values for the production cross section that lie
almost an order of magnitude below the measured values. Here, $\alpha_s$
is the strong coupling and $v$ is the velocity of the charm quark ($c$)
or charm antiquark ($\bar c$) in the charmonium rest frame.

Calculations at next-to-leading order (NLO) in
$\alpha_s$~\cite{Zhang:2005cha,Gong:2007db} and NLO in $v^2$
\cite{Bodwin:2006dn,Bodwin:2006ke,Bodwin:2007ga} seem to resolve this
discrepancy. However, the large NLO corrections raise issues about the
convergence of the $\alpha_s$ and $v$ expansions. In the case of
the $v$ expansion, the large NLO corrections arise from several sources,
each of which contributes a modest correction that is consistent with a
convergent expansion \cite{Bodwin:2006dn,Bodwin:2006ke,Bodwin:2007ga}.
In the case of the $\alpha_s$ expansion, the NLO contribution produces a
correction of about 100\%. It has been pointed out that the bulk of the NLO
correction at $B$-factory energies arises from double logarithms of
$Q^2/m_c^2$, where $Q^2$ is the square of the
$e^+e^-$-center-of-momentum energy and $m_c$ is the charm-quark mass
\cite{Jia:2010fw}. This suggests that one might gain control of the
$\alpha_s$ expansion by resumming the large double (and single)
logarithms of $Q^2/m_c^2$.

The standard tool for the resummation of logarithms in exclusive processes
is the light-cone formalism \cite{Lepage:1980fj,Chernyak:1983ej}.
However, the evolution of light-cone distributions of mesons produces
only single logarithms and, so, does not control the double
logarithms that appear in NLO amplitude for $e^+e^-\to J/\psi +\eta_c$. 
It has been suggested in Ref.~\cite{Jia:2010fw} that the
double logarithms could be related to end-point singularities in the
light-cone hard-scattering kernels.\footnote{Such singularities are
discussed in Refs.~\cite{Lepage:1980fj,Chernyak:1983ej}.} Our 
analysis confirms this conjecture.

In this paper we identify the loop-momentum regions that give rise to
the singularities in the double logarithms of $Q^2/m_c^2$ that appear in
the limit $m_c\to 0$. (An abbreviated discussion of these double
logarithms and the associated singularities was given in
Ref.~\cite{Bodwin:2013ys}.) We find that the singularities that are
associated with the large double logarithms arise from both the Sudakov
region and the end-point region of the loop-momentum integration. The
Sudakov logarithms cancel in the sum over Feynman
diagrams.\footnote{The cancellation of Sudakov logarithms for
color-singlet mesons was noted in
Refs.~\cite{Lepage:1980fj,Chernyak:1983ej}.} Our results for the
double logarithms agree diagram by diagram with those that were
obtained in the complete NLO calculations in
Refs.~\cite{Zhang:2005cha,Gong:2007db}. A new insight that follows from
our analysis is that the end-point singularity corresponds to a
pinch-singular region of the loop-momentum integration in which a
spectator-quark line becomes either soft or soft and collinear. (Our
nomenclature is that the active-quark line is the quark line to which
the virtual photon attaches and that the spectator-quark line is the
other quark line.) This pinch singularity has a (logarithmically)
divergent power count by virtue of the fact that the process $e^+e^-\to
J/\psi+\eta_c$ proceeds through a helicity flip. The insight that the
end-point singularity corresponds to a soft-quark pinch-singular region
could have important implications for the all-orders factorization of
the infrared singularities for this process and for the resummation of the
associated logarithms.\footnote{The existing proof of factorization
for exclusive double-quarkonium production in
Refs.~\cite{Bodwin:2008nf,Bodwin:2010fi} applies only to processes that
do not flip the quark helicity and, so, is not relevant to the process
$e^+ e^- \to J/\psi + \eta_c$.}

The remainder of this paper is organized as follows. In
Sec.~\ref{sec:strategy}, we outline our strategy for computing the
double logarithms and analyzing the associated singularities. We compute
the double logarithms and identify the loop-momentum regions that give
rise to the associated singularities in Sec.~\ref{sec:double_logarithms}.
In this section, we compute both the Sudakov and the end-point 
double-logarithmic contribution
for each NLO Feynman diagram. In Sec.~\ref{sec:sudakov}, we give a
general argument for the cancellation of the Sudakov double logarithms
that is based on a soft-collinear approximation and diagrammatic Ward
identities. Section \ref{sec:end-point} contains a general analysis of the
end-point region, in which we show that the end-point singularities can
yield logarithmic divergences, but not power divergences. We summarize
our results in Sec.~\ref{sec:summary}.

\section{Strategy of the computation\label{sec:strategy}}

According to the NRQCD factorization formalism~\cite{Bodwin:1994jh},
the amplitude for the process $e^+e^-\to J/\psi + \eta_c$ can be written
as a sum of products of short-distance coefficients (SDCs) with NRQCD
long-distance matrix elements (LDMEs).
The double logarithms of $Q^2/m_c^2$ that are the focus of this paper 
appear in the single SDC that arises in the NRQCD 
factorization expression for the amplitude at LO in $v$.\footnote{
Double logarithms of $Q^2/m_c^2$ were also found in the 
order-$\alpha_s v^2$ correction to the amplitude~\cite{Dong:2012xx}.
}
This SDC can be obtained perturbatively by
comparing the full-QCD amplitude $i {\cal A} [e^+ e^- \to c\bar{c}_1
({}^3S_1) + c\bar{c}_1 ({}^1S_0)]$ with the NRQCD expression for the 
amplitude. (Here the subscripts $1$ indicate that the $c\bar c$
pairs are in color-singlet states.) Because the NRQCD LDMEs for the
$c\bar{c}$ states are insensitive to momentum scales of order $m_c$ or
larger, the double logarithms of $Q^2/m_c^2$ 
in the SDC are contained entirely in the full-QCD amplitude. Therefore, 
in our calculation, we focus on the full-QCD amplitude 
$i {\cal A} [e^+ e^- \to c\bar{c}_1 ({}^3S_1) + c\bar{c}_1 ({}^1S_0)]$.

The process $e^+ e^- \to c\bar{c}_1 ({}^3S_1) + c\bar{c}_1 ({}^1S_0)$
consists of the process $e^+ e^- \to \gamma^*$, followed by the
process $\gamma^* \to c \bar c_1 ({}^3S_1) + c\bar{c}_1 ({}^1S_0)$.
Because the process $e^+ e^- \to \gamma^*$ does not receive QCD
corrections in relative order $\alpha^0\alpha_s$, we need to consider
only the amplitude $i{\cal A}[\gamma^* \to c\bar{c}_1 ({}^3S_1) +
c\bar{c}_1 ({}^1S_0)]$ in order to compute the NLO corrections relative
to the LO amplitude. (Here, $\alpha$ is the electromagnetic coupling
constant.) In the remainder of this paper, we will denote the amplitude
$i{\cal A}[\gamma^* \to c\bar{c}_1 ({}^3S_1) + c\bar{c}_1 ({}^1S_0)]$ by
$i{\cal A}$.

The process $e^+ e^- \to J/\psi + \eta_c$ does not satisfy
quark-helicity conservation. In order to produce a helicity flip, the
amplitude $i{\cal A}$ must contain at least one numerator factor $m_c$.
It follows that the amplitude is suppressed by a factor of $m_c/Q$
relative to a helicity-conserving amplitude. In our calculation, we
retain all numerator terms that are proportional to powers of $m_c$,
except as noted. We also keep $m_c$ nonzero in denominators in order to
regulate singularities.

Our strategy in analyzing the double logarithms in $i{\cal A}$ is to
examine the double-logarithmic singularities that appear in the limit  
$m_c\to 0$. In characterizing these singularities, we ignore 
powers of $m_c$ in the coefficients of the logarithms. 
From a general Landau analysis of pinch singularities
\cite{Landau:1959fi,Coleman:1965xm,Sterman:1978bi,Sterman:1978bj,Libby:1978bx},
we expect singularities to arise from regions of loop momentum in which
an internal line is soft and/or collinear to an external line. If $m_c$
is nonzero, then there is no pinch when internal quark lines are soft or
collinear or when internal gluon lines are collinear. However, there can
still be a pinch when an internal gluon line is soft. We regulate these
soft singularities by using dimensional regularization in
$d=4-2\epsilon$ dimensions. The soft singularities then produce single
poles in $\epsilon$. The amplitudes that we consider are ultraviolet (UV)
finite. However, after we reduce tensor integrals to scalar integrals,
individual terms can contain UV divergences, which we also regulate
dimensionally and which also produce single poles in $\epsilon$. These
UV divergences cancel in the sum of the contributions from each Feynman
diagram.

Although we work in $d=4-2\epsilon$ dimensions, we evaluate Dirac traces
and numerator algebra in four dimensions. That simplification does not
affect the calculation of the double logarithms of $m_c$ because terms
of order $\epsilon$ in the numerator can contribute only in conjunction
with a soft pole or a UV pole. The coefficient of a soft pole can
contain only a single (collinear) logarithm of $m_c$, while the
coefficient of a UV pole cannot contain any logarithm of $m_c$.
Following this procedure for the numerator algebra, we reproduce all
of the double logarithms of $m_c$ that appear in the exact NLO
calculation of Refs.~\cite{Zhang:2005cha,Gong:2007db}.

We will see that the double logarithmic singularities that appear
in the limit $m_c\to 0$ arise from two sources. One source is a
region of loop momentum in which the momentum of a gluon becomes both soft
and collinear. This is the ``Sudakov'' region. A second source is
a region of loop momentum in which one gluon carries away almost all of
the momentum of a spectator quark and the other gluon carries away almost
all of the momentum of a spectator antiquark. This is the ``end-point''
region. As we will show, the end-point region can also be characterized
as a momentum region in which a spectator-quark line becomes either soft
or soft and collinear. In a general analysis of pinch singularities,
such a momentum configuration produces a pinch. Furthermore, as is
evident from our calculation, in order $m_c/Q$, the pinch has the
correct power counting to produce a (logarithmic) singularity. In
contrast, in order $(m_c/Q)^0$, this same pinch does not have the
correct power counting to produce a singularity. Consequently, end-point
momentum configurations do not play a role in non-helicity-flip
processes at the leading nontrivial order in $m_c/Q$.\footnote{The
suppression of end-point singularities by inverse powers of the large
momentum transfer was noted in
Refs.~\cite{Drell:1969km,West:1970av,Lepage:1980fj,Chernyak:1983ej}.}

\section{Calculation of the double logarithms}
\label{sec:double_logarithms}%

In this section, we evaluate the double logarithms that appear in the NLO
QCD corrections to the amplitude $\gamma^* \to J/\psi+\eta_c$, and we
identify the momentum regions that are associated with the 
double-logarithmic singularities in the limit $m_c\to 0$.

\subsection{Kinematics, conventions, and nomenclature}

First, we describe the kinematics, conventions, and nomenclature
that we use in calculating the double logarithms and throughout this
paper. We work in the Feynman gauge. We use the light-cone momentum
coordinates $k=[k^+,k^-,\bm{k}_\perp]=
[(k^0+k^3)/\sqrt{2},(k^0-k^3)/\sqrt{2},\bm{k}_\perp]$ and work in the
$e^+e^-$-center-of-momentum frame. Because our calculation is at LO in
$v$, we set the relative momentum of the $c$ and $\bar c$ in each
charmonium equal to zero. Then, the momenta of the $c$ and $\bar{c}$ in
the $J/\psi$ are both
$p=[(\sqrt{P^2+m_c^2}+P)/\sqrt{2},(\sqrt{P^2+m_c^2}-P)/\sqrt{2},\bm{0}_\perp]$,
and the momenta of the $c$ and $\bar{c}$ in the $\eta_c$ are both $\bar
p=[(\sqrt{P^2+m_c^2}-P)/\sqrt{2},(\sqrt{P^2+m_c^2}+P)/\sqrt{2},\bm{0}_\perp]$,
where $P$ is the magnitude of the 3-momentum of any of the $c$'s or
$\bar c$'s. The momentum of the virtual photon is $2(p+\bar p)$, which
implies that $Q^2=16(P^2+m_c^2)$. Note that $p^+ = \bar p^-\sim Q$ and
$p^-=\bar p^+\sim m_c^2/Q$.

If a momentum $k$ has light-cone components whose orders of magnitude
are $P\lambda[1,(\eta^+)^2,\eta^+]$, then we say that $k$ is soft if
$\lambda\ll 1$, and we say that $k$ is collinear to plus if $\eta^+\ll
1$. If $k$ has light-cone components whose orders of magnitude are
$P\lambda[(\eta^-)^2,1,\eta^-]$, then we say that $k$ is soft if
$\lambda\ll 1$, and we say that $k$ is collinear to minus if $\eta^-\ll
1$. Hence, $p$ is collinear to plus and $\bar p$ is collinear to minus
in the limit $m_c^2/P^2\to 0$.

The amplitudes that we compute contain spin and
color projectors that put the $Q\bar Q$ pairs into states of definite
spin and color~\cite{Bodwin:2002hg}. When the relative momentum of the
$c$ and $\bar c$ in each charmonium is zero, the
$\hbox{spin-singlet}\otimes \hbox{color-singlet}$ and
$\hbox{spin-triplet}\otimes \hbox{color-singlet}$ projectors are given
by
\begin{subequations}
\begin{eqnarray}
&&
\Pi_1 (\bar p, \bar p) =- \frac{1}{2 \sqrt{2} m_{c}} 
\gamma^5 (\bar p \sl +
m_c) \otimes \frac{{\bf 1}}{\sqrt{N_c}}
=- \frac{1}{2 \sqrt{2} m_{c}} (-\bar p \sl +
m_c)\gamma^5  \otimes \frac{{\bf 1}}{\sqrt{N_c}},\\
&&
\Pi_3 (p,p,\lambda) =- \frac{1}{2 \sqrt{2} m_{c}} 
\epsilon \sl^* (\lambda)
(p\sl + m_c) \otimes \frac{{\bf 1}}{\sqrt{N_c}}
=- \frac{1}{2 \sqrt{2} m_{c}} 
(-p\sl + m_c) \epsilon \sl^* (\lambda)\otimes \frac{{\bf 1}}{\sqrt{N_c}},
\phantom{xxx}
\end{eqnarray}
\end{subequations}
where $\epsilon^* (\lambda)$ is the polarization vector for the 
$c\bar c$ pair in the spin-triplet state, $N_c=3$ is the number of
colors, ${\bf 1}$ is the unit color matrix, and we use 
nonrelativistic normalization for the spinors.

\subsection{Evaluation of the LO diagrams}

\begin{figure}                                                   
\epsfig{file=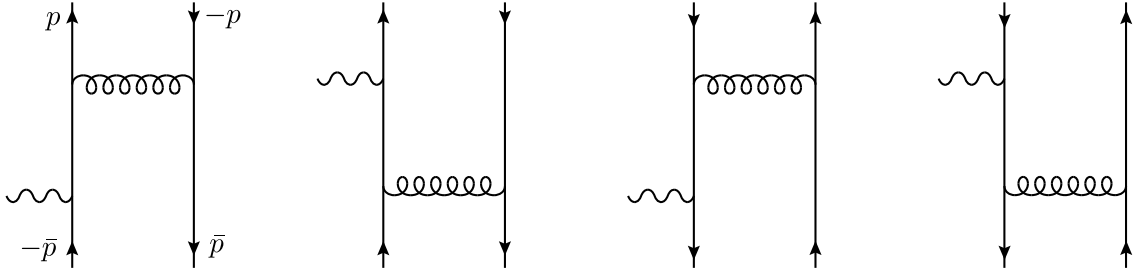,width=15.0cm}
\caption{Feynman diagrams for the process $\gamma^*\to J/\psi+\eta_c$ at
LO in $\alpha_s$. The upper $c\bar c$ pair corresponds to the $J/\psi$,
and the lower $c\bar c$ pair corresponds to the $\eta_c$.
\label{fig:LO}
}
\end{figure}

In Fig.~\ref{fig:LO}, we show the diagrams at LO in
$\alpha_s$ for the process $\gamma^*\to J/\psi+\eta_c$. A              
straightforward computation of the contribution to the amplitude 
from  these diagrams yields                                     
\begin{equation}                                                   
i {\cal A}_{\rm LO}^{\mu} = \frac{-i 256 \pi \alpha_s C_F}{m_c Q^4}
\epsilon^{\mu \nu \alpha \beta}
\epsilon^*_\nu (\lambda) p_\alpha \bar p_\beta,
\label{LO-amp}
\end{equation}
where $C_F = (N_c^2-1)/(2N_c)$ and $\epsilon^{\mu \nu \alpha \beta}$ is
the totally antisymmetric tensor in four dimensions, for which we use the
convention $\epsilon_{0123} = +1$. We have suppressed the factor $-i$
times the charm-quark charge that is associated with the electromagnetic
vertex.
\subsection{Evaluation of the double logarithms in the NLO diagrams}
\label{sec:evaluation}%

\begin{figure}
\epsfig{file=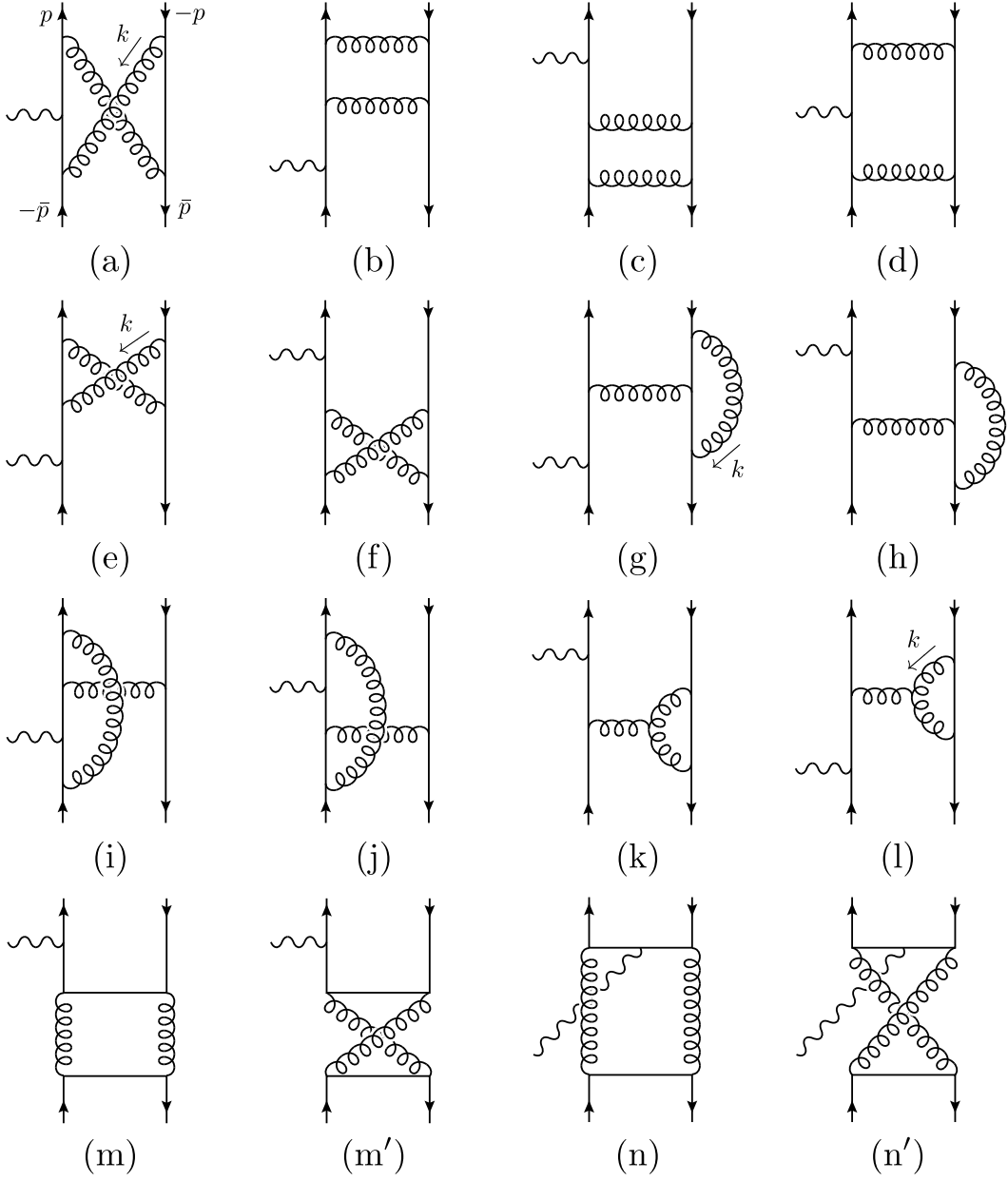,width=14cm}
\caption{One-loop diagrams for the process $e^+e^-\to J/\psi + \eta_c$
that potentially contain double logarithms in $Q^2/m_c^2$. The
upper $c\bar c$ pair corresponds to the $J/\psi$, and the lower $c\bar
c$ pair corresponds to the $\eta_c$. We do not show diagrams that are
charge conjugates of the diagrams in the figure.
\label{fig:oneloop}
}
\end{figure}

Now we calculate the double logarithms that arise from the Feynman
diagrams that contribute to the NLO QCD corrections to the amplitude.
As we will explain in Secs.~\ref{sec:ident-sudakov} and
\ref{sec:ident-end-point}, only certain diagrams can potentially yield
double logarithms of $Q^2/m_c^2$. These diagrams are shown in
Fig.~\ref{fig:oneloop}.

\begin{figure}
\epsfig{file=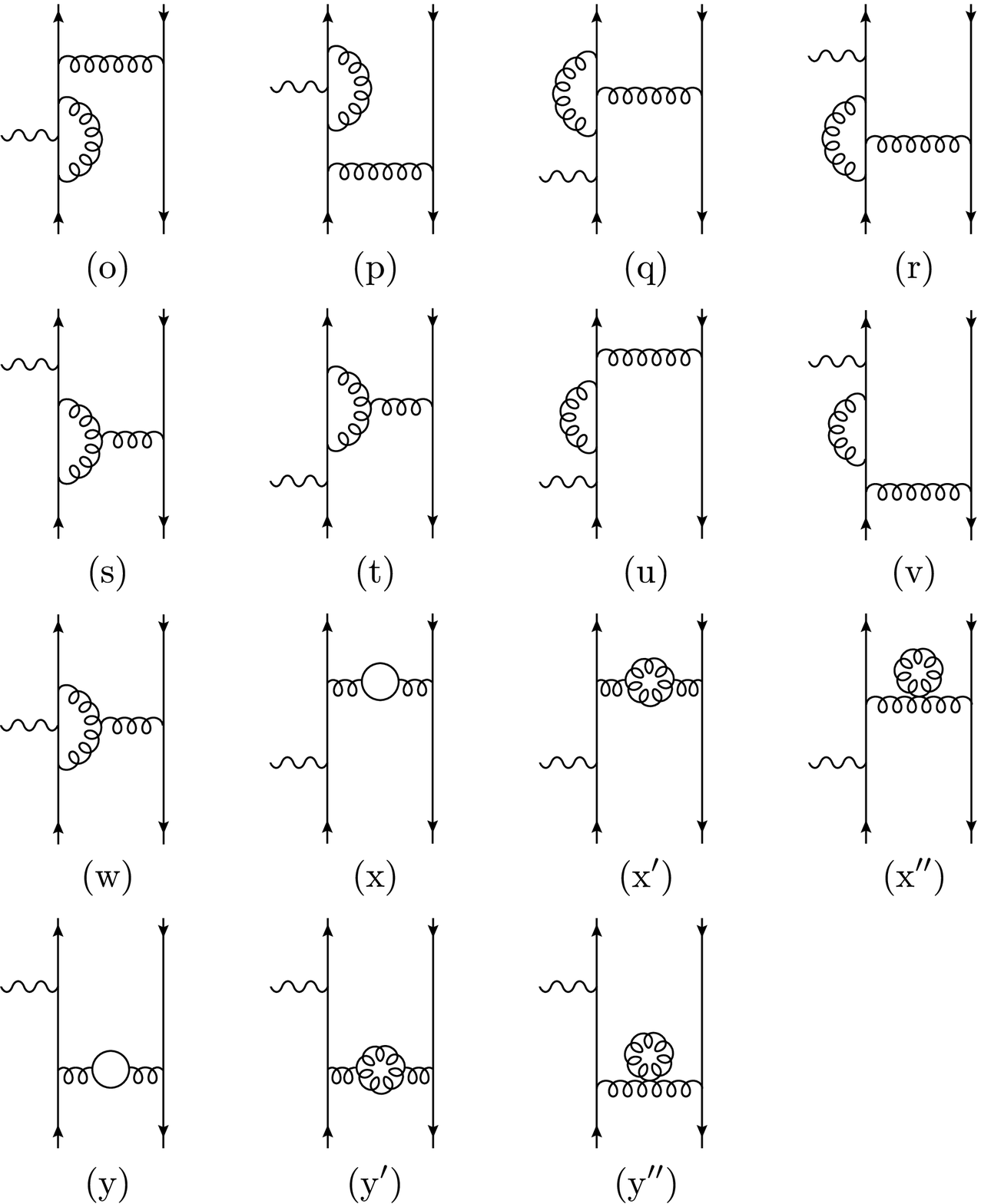,width=14cm}
\caption{One-loop diagrams for the process $e^+e^-\to J/\psi +
\eta_c$ that do not contain the double logarithms in $Q^2/m_c^2$.
The upper $c\bar c$ pair corresponds to the $J/\psi$, and the
lower $c\bar c$ pair corresponds to the $\eta_c$. We do not show
diagrams that are charge conjugates of the diagrams in the figure.
\label{fig:oneloop_nodl}
}
\end{figure}

\subsubsection{Diagram of Figure~\ref{fig:oneloop}(a)}

In order to illustrate our methods, we discuss in some detail the diagram
that is shown in Fig.~\ref{fig:oneloop}(a). The associated amplitude is
\begin{eqnarray}
i {\cal A}_{\rm NLO}^{\rm (a)\mu} &=& 
\int_k \frac{1}{N_c}{\rm Tr} \bigg[ \Pi_3(p,p,\lambda)
(-i g_s \gamma^\alpha T^a) 
\frac{i (k \sl + 2 p \sl + \bar p \sl+m_c)}
{(k+2 p+\bar p)^2 -m_c^2 +i \varepsilon} \gamma^\mu
\frac{i (k \sl - \bar p \sl+m_c)}
{(k-\bar p)^2 -m_c^2 +i \varepsilon} 
\nonumber \\ && 
\times
(-i g_s \gamma^\beta T^b) 
\Pi_1 (\bar p,\bar p) (-i g_s \gamma_\alpha T^a) 
\frac{i (-k \sl - p \sl+m_c)}
{(k+p)^2 -m_c^2 +i \varepsilon} (-i g_s \gamma_\beta T^b) \bigg]
\nonumber \\ && 
\times 
\frac{-i}{k^2 + i \varepsilon}
\frac{-i}{(k+p+\bar p)^2 + i \varepsilon},
\label{decomp}
\end{eqnarray}
where $g_s = \sqrt{4 \pi \alpha_s}$, $T^a$ is the generator of the
fundamental (triplet) representation of $SU(3)$ with 
adjoint-representation color index $a\in \{1,\,2,\,\ldots,\,N_c^2-1\}$, and
the trace is over the gamma matrices and color matrices. The symbol $\int_k$ is
defined by
\begin{equation}
\int_k  \equiv \mu^{2 \epsilon} \int \frac{d^dk}{(4 \pi)^d},
\end{equation}
where $\mu$ is the dimensional-regularization scale. As we have
mentioned, because the process $e^+ e^- \to J/\psi + \eta_c$ does not
satisfy quark-helicity conservation, we retain factors of $m_c$ in the
numerator. Factors of $m_c$ can come from the numerators of the quark
propagators or from the numerators of the spin projectors.

As a first step in reducing the tensor integrals in Eq.~(\ref{decomp}) to 
scalar integrals, we decompose the loop momentum $k$ as follows:
\begin{equation}
\label{eq:decomposition}%
k = 
\frac{k \cdot (p+\bar p)}{(p + \bar p)^2} 
(p+ \bar p) 
+ 
\frac{k \cdot (p-\bar p)}{(p - \bar p)^2} 
(p- \bar p) 
+ k_\perp, 
\end{equation}
which is valid to all orders in $m_{c}^2/Q^2$. Here,
$k_\perp=[0^+,0^-,\bm{k}_\perp]$. Because the propagator
denominators and the numerator factors $k\cdot p$ and $k\cdot \bar{p}$
are independent of the angles of $\bm{k}_\perp$, 
we can carry out the average over the
angles of $\bm{k}_\perp$ easily, setting terms that are linear or cubic in
$k_\perp$ to zero and making the replacement $k_\perp^\mu
k_\perp^\nu\to -(1/2)g^{\mu\nu}_{\perp}\bm{k}_\perp^2 $ for terms that
are quadratic in $k_\perp$, where $-g_{\perp}^{\mu\nu}$ can be
expressed as
\begin{equation}
-g_{\perp}^{\mu\nu}
=
-g^{\mu\nu}
+\frac{(p+\bar{p})^\mu(p+\bar{p})^\nu}{(p+\bar{p})^2}
+\frac{(p-\bar{p})^\mu(p-\bar{p})^\nu}{(p-\bar{p})^2}.
\end{equation}
After carrying out the
trace over the gamma and color matrices, we find that
$i {\cal A}_{\rm NLO}^{\rm (a)\mu}$ can be written as
\begin{eqnarray}
\label{eq:diaga}%
i {\cal A}_{\rm NLO}^{\rm (a)\mu} &=& 
\frac{2 g_s^4}{m_c} C_F \left( C_F - \frac{C_A}{2}  \right) 
\epsilon^{\mu \nu \alpha \beta}
 \epsilon^*_\nu (\lambda) p_\alpha \bar p_\beta
\int_k 
\frac{1}{\cal D} \bigg[
3 k^2 + 4 k \cdot p - 2 k \cdot \bar p - 4 p \cdot \bar p
 +\bm{k}_\perp^2 
\nonumber \\ && 
+ \frac{4 (p \cdot \bar p k \cdot p - m_c^2 k \cdot \bar p)}{(p+\bar p)^2
(p-\bar p)^2} 
(k^2 + 4 k \cdot p + 2 k \cdot \bar p)  
 \bigg],
\end{eqnarray}
where $C_A = N_c$, ${\cal D} = D_0 D_1 D_2 D_3 D_4$, and 
\begin{subequations}
\begin{eqnarray}
D_0 &=& k^2 +i \varepsilon, 
\\
D_1 &=& (k-\bar p)^2 - m_c^2 +i \varepsilon, 
\\
D_2 &=& (k+p)^2 - m_c^2 +i \varepsilon,
\\
D_3 &=& (k+p+\bar p)^2 +i \varepsilon, 
\\
D_4 &=&  (k+2 p + \bar p)^2 - m_c^2 +i \varepsilon.
\end{eqnarray}
\end{subequations}
We can then use the identity 
\begin{equation}
\bm{k}_\perp^2 = 
\frac{[k \cdot (p+\bar p)]^2}{(p+\bar p)^2} 
+ 
\frac{[k \cdot (p-\bar p)]^2}{(p-\bar p)^2} 
-k^2, 
\end{equation}
to eliminate $\bm{k}_\perp^2$. Writing the numerator in terms of the 
$D_i$'s, we obtain
\begin{eqnarray}
i {\cal A}_{\rm NLO}^{\rm (a)\mu} &=& 
\frac{2 g_s^4}{m_c} C_F \left( C_F - \frac{C_A}{2} \right) 
\epsilon^{\mu \nu \alpha \beta}
 \epsilon^*_\nu (\lambda) p_\alpha \bar p_\beta
\int_k \frac{1}{\cal D} 
\bigg[
\frac{D_0 D_3}{(p-\bar p)^2} 
- 
\frac{3 D_2 D_3}{(p - \bar p)^2} 
+ \frac{D_0 D_2}{(p + \bar p)^2} 
\nonumber \\ && 
- 
\frac{2 p \cdot \bar p}{(p + \bar p)^4}
\left( D_0 D_1 + D_3 D_4 \right)
+ \frac{8m_c^2 p \cdot \bar p
D_1 D_3 
}{(p -\bar p)^2 (p + \bar p)^4} 
+ \frac{6 (p\cdot \bar p)^2 + 21 m_c^2 p\cdot \bar p + 7 m_c^4 }
{(p -\bar p)^2 (p + \bar p)^4} D_2 D_4
\nonumber \\ && 
- \frac{4 [m_c^4 + (p \cdot \bar p)^2] D_0 D_4}
{(p -\bar p)^2 (p + \bar p)^4} 
- \frac{2 (p\cdot \bar p)^2 - 5 m_c^2 p\cdot \bar p + m_c^4 }
{(p -\bar p)^2 (p + \bar p)^4} 
\left( D_1 D_2 - 2 D_1 D_4 
\right)
\nonumber \\ && 
+ \left( 6 - \frac{2 m_c^2 ( 9 m_c^2 + p \cdot \bar p)}
{(p -\bar p)^2 (p + \bar p)^2} 
\right) D_2
\bigg],
\end{eqnarray}
where we have made use of the identities
\begin{subequations}
\begin{eqnarray}
&&2D_0-D_1-2D_3+D_4=0,\\
&&D_1-2D_2+D_4=4(p\cdot\bar{p}+m^2_c).
\end{eqnarray}
\end{subequations}

The scalar integrals can be evaluated by using standard methods. 
If we ignore all integrals that do not produce the double logarithms in
$Q^2/m_c^2$, we obtain
\begin{eqnarray}
\label{diagram_a}%
i {\cal A}_{\rm NLO}^{\rm (a)\mu} &\approx& 
i {\cal A}_{\rm LO}^{\mu} 
\frac{-i \alpha_s \pi Q^2}{2} 
\left( C_F - \frac{C_A}{2} \right) 
\nonumber \\ 
&& 
\times \bigg \{
\int_k \frac{1}{(k^2+i \varepsilon) 
[ (k+p)^2 - m_c^2 + i \varepsilon] 
[ (k-\bar p)^2 -m_c^2 +i \varepsilon] }
\nonumber \\ && \hspace{1.5ex}
+ 
\int_k \frac{1}{[(k+p+\bar p)^2+i \varepsilon] 
[ (k+2 p+\bar p)^2 - m_c^2 + i \varepsilon] 
[ (k+p)^2 -m_c^2 +i \varepsilon] }
\nonumber \\ && \hspace{1.5ex}
+
\int_k \frac{1}{(k^2+i \varepsilon) 
[ (k+p)^2 - m_c^2 + i \varepsilon] 
[ (k+p+\bar p)^2 +i \varepsilon] }
\bigg \},
\end{eqnarray}
where the LO amplitude is given in Eq.~(\ref{LO-amp}).
\subsubsection{The scalar integral ${\cal S}$}

Let us consider the first scalar integral in Eq.~(\ref{diagram_a}),
which we denote by ${\cal S}$:
\begin{eqnarray}
{\cal S} 
&\equiv& 
\int_k \frac{1}{(k^2+i \varepsilon) 
[ (k+p)^2 - m_c^2 + i \varepsilon] 
[ (k-\bar p)^2 -m_c^2 +i \varepsilon] }
\nonumber \\ 
&=& 
\int_{\bm{k}_\perp} 
\int \frac{dk^+}{2 \pi} 
\int \frac{dk^-}{2 \pi} 
 \frac{1}{(2 k^+ k^- - \bm{k}_\perp^2 + i \varepsilon)
( 2 k^+ k^- - \bm{k}_\perp^2 + 2 k^+ p^- + 2 k^- p^+ + i \varepsilon) }
\nonumber \\ 
&& 
\times 
\frac{1}{ 
2 k^+ k^- - \bm{k}_\perp^2 - 2 k^+ \bar p^- - 2 k^- \bar p^+ + i \varepsilon 
},
\label{cal-S} 
\end{eqnarray}
where 
\begin{equation}
\int_{\bm{k}_\perp}\equiv\mu^{2\epsilon}\int 
\frac{d^{d-2}\bm{k}_\perp}{(2\pi)^{d-2}}.
\end{equation}
We remind the reader that, because we work in
the center-of-momentum frame, $p^+ = \bar p^-\sim Q$ and $p^-=\bar
p^+\sim m_c^2/Q$.

As we have explained, we can understand the origin of the double
logarithms of $Q^2/m_c^2$ in $\cal{S}$ by analyzing the pinch
singularities in $\cal{S}$. From a general Landau analysis
\cite{Landau:1959fi,Coleman:1965xm,Sterman:1978bi,Sterman:1978bj,Libby:1978bx},
we know that the only possible pinch singularities are those that
correspond to $k$ soft, collinear to plus, and collinear to minus
\cite{Sterman:2004pd}. We now verify by direct examination of $\cal{S}$
that those pinch singularities are present.

In the $k^-$ complex plane, the integrand in Eq.~(\ref{cal-S}) has
three poles, which are located at
\begin{subequations}
\label{eq:sudakov_kminus_poles}
\begin{eqnarray}
&& 
k^- = \frac{\bm{k}_\perp^2 -i \varepsilon}{2 k^+}, 
\\
&& 
k^- = \frac{\bm{k}_\perp^2 - 2 k^+ p^--i \varepsilon}{2 (k^++p^+)},
\\
&& 
k^- = \frac{\bm{k}_\perp^2 + 2 k^+ \bar p^- -i \varepsilon}
{2 (k^+ - \bar p^+)}.
\end{eqnarray}
\end{subequations}
In the $k^+$ complex plane, the integrand in Eq.~(\ref{cal-S}) has
three poles, which are located at
\begin{subequations}
\label{eq:sudakov_kplus_poles}
\begin{eqnarray}
&& 
k^+ = \frac{\bm{k}_\perp^2 -i \varepsilon}{2 k^-}, 
\\
&& 
k^+ = \frac{\bm{k}_\perp^2 - 2 k^- p^+-i \varepsilon}{2 (k^-+p^-)},
\\
&& 
k^+ = \frac{\bm{k}_\perp^2 + 2 k^- \bar p^+ -i \varepsilon}
{2 (k^- - \bar p^-)}.
\end{eqnarray}
\end{subequations}
In the $k_{\perp}^i$ complex plane, the integrand in Eq.~(\ref{cal-S}) has 
six poles, which are located at 
\begin{subequations}
\label{eq:sudakov_kperp_poles}
\begin{eqnarray}
&& 
k_{\perp}^i=\pm\sqrt{2k^+k^--(k_{\perp}^j)^2+i\varepsilon},
\\
&& 
k_{\perp}^i=\pm\sqrt{2(k^+k^-+k^+p^-+k^-p^+)-(k_{\perp}^j)^2+i\varepsilon},
\\
&& 
k_{\perp}^i=\pm\sqrt{2(k^+k^--k^+\bar p^--k^-\bar p^+)-(k_{\perp}^j)^2
+i\varepsilon},
\end{eqnarray}
\end{subequations}
where $(k_{\perp}^j)^2$ is summed over $j\neq i$ in the dimensionally 
regulated expression.

Consider first the case $m_c\neq 0$. The poles in $k^-$ pinch the $k^-$
contour at $k^-=0$ when $k^+$ and $\bm{k}_\perp$ go to zero in a fixed
ratio. The combinations of poles that provide a pinch depend on the sign
of $k^+$. Similarly, the $k^+$ contour is pinched at $k^+=0$ when $k^-$
and $\bm{k}_\perp$ go to zero in a fixed ratio. The $k_{\perp}^i$
contour is also pinched at $k_{\perp}^i=0$ when $k^+$, $k^-$,
and $k_{\perp}^j$ go to zero. These pinches correspond to the soft
singularity at $k=0$.\footnote{It is possible that some, but not all, of
the components of $k$ can have pinched contours of integration. In this
case, it may be possible to avoid the singular region by deforming the
unpinched contour. For example, the second and third denominators in
Eq.~(\ref{cal-S}) pinch the $k^-$ contour at $k^-=-p^- + \bar p^-$ and
pinch the $k^+$ contour at $k^+=-p^+ + \bar p^+$. However, it is easy to
see that, for these values of $k^-$ and $k^+$,  the $k_{\perp}^i$
contour is not pinched at $k_{\perp}^i=0$, even if
$k_{\perp}^j=0$. It can also be seen that, for these values of $k^-$
and $k^+$, the first denominator in Eq.~(\ref{cal-S}) is off shell by
order $p^+\bar p^-\sim Q^2$, and so is inconsistent with the scaling for
a soft singularity.}

Now consider the case $m_c=0$. Again there are pinches corresponding to
a soft singularity: the $k^-$ contour is pinched at
$k^-=0$ when $k^+$ and $\bm{k}_\perp$ go to zero in a fixed ratio; the
$k^+$ contour is pinched at $k^+=0$ when $k^-$ and $\bm{k}_\perp$ go to
zero in a fixed ratio; and the $k_{\perp}^i$ contour is pinched at
$k_{\perp}^i=0$ when $k^+$, $k^-$, and $k_{\perp}^j$ go to zero.
(The combinations of poles that provide the pinches for given signs of
$k^+$ and $k^-$ are different in the massless case than in the massive
case.)

In the case $m_c=0$, there are additional pinches that correspond to
collinear singularities. The first and second $k^-$ poles pinch the
$k^-$ contour at $k^- = 0$ when $-p^+ < k^+ <0$ and $\bm{k}_\perp$ goes
to zero such that $\bm{k}_\perp^2/k^+=0$ and
$\bm{k}_\perp^2/(k^++p^+)=0$. The $k_{\perp}^i$ contour is also pinched
at $k_{\perp}^i=0$ when $k^-$ and $k_{\perp}^j$ go to zero. These pinches
correspond to a collinear-to-plus singularity. Note that the pinches in
the $k^-$ and $k_{\perp}^i$ contours occur even when $k^+$ is
arbitrarily close to $0$ or to $ -p^+$. There is no pinch in the $k^+$
contour, but this is consistent with a collinear-to-plus momentum
configuration, in which $k^+$ takes on any value in the range $-p^+ <
k^+ <0$. When $m_c=0$, there are also pinches that correspond to a
collinear-to-minus singularity: the first and third $k^+$ poles pinch
the $k^+$ contour at $k^+ = 0$ when $0<k^-<\bar p^-$ and $\bm{k}_\perp$
goes to zero such that $\bm{k}_\perp^2/k^-$ goes to zero and
$\bm{k}_\perp^2/(k^--\bar p^-)$ goes to zero; 
the $k_{\perp}^i$ contour is pinched
at $k_{\perp}^i=0$ when $k^+$ and $k_{\perp}^j$ go to zero.

We conclude that there are pinches in ${\cal S}$ that correspond
to soft, collinear-to-plus, and collinear-to-minus singularities, as
expected. These pinches correspond to Sudakov logarithms of $m_c$.
In particular, the double logarithms of $m_c$ result from an overlap
between the soft and collinear pinches \cite{Sterman:2004pd}.

${\cal S}$ can be evaluated easily by combining denominators with
Feynman parameters. The result is
\begin{eqnarray}
{\cal S} &=& 
\frac{i}{4 \pi^2 Q^2}\,
\frac{p^++p^-}{p^+-p^-} 
\bigg[ \left( \frac{1}{\epsilon_{\rm IR}} + \log \frac{4 \pi \mu^2
e^{-\gamma_{{}_{\rm E}}}}{m_c^2} 
\right)
\left( \log \frac{p^-}{p^+} + i \pi \right)
\nonumber \\ && 
+ 2\; {\rm Sp} \left( 1-\frac{p^-}{p^+} \right) 
+ \frac{1}{2} \log^2\frac{p^-}{p^+}
- \pi^2 + i \pi \log \frac{p^+ p^-}{(p^+-p^-)^2}
+ O(\epsilon)
\bigg],
\end{eqnarray}
where $\gamma_{{}_{\rm E}}$ is the Euler{-}Mascheroni constant, and
${\rm Sp}(z)$ is the Spence function:
\begin{equation}
{\rm Sp}(z)=-\int_0^z du \frac{\log(1-u)}{u},
\end{equation}
for any complex number $z \notin [1,\infty)$. The subscript IR on
$\epsilon$ indicates that the pole is associated with the infrared
divergence. If we retain only the terms that are singular in $\epsilon$
and the double logarithms, then we have
\begin{equation}
{\cal S} \approx
\frac{i}{4 \pi^2 Q^2} 
\bigg[ \left( -\frac{1}{\epsilon_{\rm IR}} + \log \frac{m_c^2}{\mu^2}
\right)
\log \frac{Q^2}{m_c^2} + \frac{1}{2} \log^2 \frac{Q^2}{m_c^2} \bigg].
\end{equation}
The single logarithm of $Q^2/m_c^2$ corresponds, in the limit $m_c\to
0$, to the collinear-to-plus and collinear-to-minus singularities. The
double logarithm of $Q^2/m_c^2$ corresponds, in the limit $m_c\to
0$, to the overlap of the soft singularity and the collinear-to-plus and
collinear-to-minus singularities. The single logarithm of $m_c^2/\mu^2$
corresponds to the soft singularity.
 
In the second integral in Eq.~(\ref{diagram_a}), we can write
$k'=k+p+\bar p$, where $k+p+\bar p$ is the momentum of the gluon that
connects the quark line with momentum $p$ and the quark line with
momentum $\bar p$. Then, this integral becomes
\begin{eqnarray}
&& \hspace{-5ex} 
\int_k \frac{1}{[(k+p+\bar p)^2+i \varepsilon] 
[ (k+2 p+\bar p)^2 - m_c^2 + i \varepsilon] 
[ (k+p)^2 -m_c^2 +i \varepsilon] }
\nonumber \\ &=& 
\int_{k'} \frac{1}{(k'^2+i \varepsilon) 
[ (k'+p)^2 - m_c^2 + i \varepsilon] 
[ (k'-\bar p)^2 -m_c^2 +i \varepsilon] }, 
\end{eqnarray}
which is identical to $\cal S$. 

\subsubsection{The scalar integral ${\cal E}$}

Now we consider the third integral in Eq.~(\ref{diagram_a}), which 
we denote by ${\cal E}$:
\begin{equation}
{\cal E} \equiv 
\int_k \frac{1}{(k^2 + i \varepsilon)
[ (k+p)^2 - m_c^2 + i \varepsilon] 
[ (k+p+\bar p)^2 + i \varepsilon] }.
\label{cal-E}
\end{equation}
If we change the loop momentum to the spectator-quark momentum $\ell =
-k -p$, then $\cal E$ can be written as
\begin{equation}
{\cal E} 
= 
\int_\ell \frac{1}{
( \ell^2 - m_c^2 + i \varepsilon) 
[ (\ell+p)^2 + i \varepsilon] 
[ (\ell-\bar p)^2 +i \varepsilon] 
}. 
\label{reinterpret}
\end{equation}
Except for the $m_c^2$ terms in the denominator factors, this expression
is identical to the one for ${\cal S}$ in Eq.~(\ref{cal-S}). Hence, in
the limit $m_c\to 0$, ${\cal E}$ will develop singularities when $\ell$,
the spectator-quark momentum, becomes soft and/or collinear to plus or
collinear to minus. In ${\cal S}$, the collinear singularities were
regulated by $m_c$, while the soft singularities required an additional
regulator, which we took to be dimensional. In contrast, as can be seen
by inspection, both the soft and collinear singularities in ${\cal E}$
are regulated by $m_c$.

When $\ell$ is soft, the gluon with momentum $k$ carries away all of
the collinear-to-plus momentum of the spectator quark with momentum
$-p$, and the gluon with momentum $k+p+\bar p$ carries away all of the
collinear-to-minus momentum of the spectator quark with momentum $\bar
p$. Therefore, the region of loop momentum in which $\ell$ is soft
corresponds to the end-point region. From our analysis of ${\cal S}$, we
know that the double logarithms arise from contributions in which $\ell$
is simultaneously soft and collinear. Therefore, the double logarithms
in ${\cal E}$ are end-point double logarithms.

${\cal E}$ can be evaluated straightforwardly by using Feynman
parameters to combine denominators. The result is
\begin{equation}
{\cal E} = 
 \frac{i}{4 \pi^2 Q^2}\,
\frac{p^+ + p^-}{p^+ - p^-} 
\bigg[ 
\frac{1}{2} \log^2 \frac{p^-}{p^+}
+ 2 \; {\rm Sp} \left( - \frac{p^-}{p^+} \right)
+ \frac{\pi^2}{6} + i \pi \log \frac{p^-}{p^+} 
+O(\epsilon)
\bigg] . 
\end{equation}
If we expand the result in powers of $m^2/Q^2$ and retain only the double
logarithms in $Q^2/m_c^2$, then we obtain
\begin{equation}
{\cal E} \approx \frac{i}{8 \pi^2 Q^2} \log^2 \frac{Q^2}{m_c^2}.
\end{equation}

\subsubsection{Summary of the double logarithms}
Direct evaluation of the remaining diagrams in Fig.~\ref{fig:oneloop}
shows that the double logarithms in each diagram arise solely from the
two scalar integrals $\cal S$ and $\cal E$. A summary of the double
logarithms from each diagram is given in
Table~\ref{Table:Double_Logarithms}. The authors of
Ref.~\cite{Jia:2010fw} have also computed these double logarithms, and
our results agree diagram by diagram with their results
\cite{jia-private-comm}.

We note that the Sudakov double logarithms cancel in the sum over all
diagrams, and, so, the double logarithm in the NLO correction to the
amplitude is given entirely by the sum of the end-point double
logarithms. Multiplying by a factor of two to take into account the
charge conjugates of the diagrams in Fig.~\ref{fig:oneloop}, we obtain
for the double logarithm in the NLO correction
\begin{eqnarray}
i {\cal A}_{\rm NLO}^\mu &\approx& 
i {\cal A}_{\rm LO}^\mu 
(-i \alpha_s \pi Q^2) 
\left[ \frac{7}{2} C_F + ( 2 C_F - C_A )\right]
{\cal E}
\nonumber\\ 
&\approx& 
i {\cal A}_{\rm LO}^{\mu} 
\,\frac{7 N_c^2 -11}{32 N_c}\,
\frac{\alpha_s}{\pi} 
\log^2 \frac{Q^2}{m_c^2},
\end{eqnarray}
in agreement with the results in Refs.~\cite{Gong:2007db,Jia:2010fw}.

\begin{table}[t]
\caption{
\label{Table:Double_Logarithms}%
End-point and Sudakov double logarithms that arise from each diagram
in Fig.~\ref{fig:oneloop} in units of $i {\cal A}_{\rm LO} \times
(-i \alpha_s \pi Q^2)/2$.
}
\begin{ruledtabular}
\begin{tabular}{c|cc}
Diagram & End-point double logarithm & Sudakov double logarithm
\\
\hline
(a) & $(C_F-\frac{1}{2} C_A) {\cal E}$ & $2 (C_F-\frac{1}{2} C_A) {\cal S}$
\\
(b) & $\phantom{1}\; C_F {\cal E}$ & $0$
\\
(c) & $2\; C_F {\cal E}$ & $0$
\\
(d) & $\frac{1}{2}\; C_F {\cal E}$ & $0$
\\
(e) & $(C_F - \frac{1}{2} C_A){\cal E}$ &
$\phantom{-} (C_F - \frac{1}{2} C_A) {\cal S}$
\\
(f) & $0$ & $\phantom{-}(C_F - \frac{1}{2} C_A) {\cal S}$
\\
(g) & $0$ & $-(C_F - \frac{1}{2} C_A) {\cal S}$
\\
(h) & $0$ & $-(C_F - \frac{1}{2} C_A) {\cal S}$
\\
(i) & $0$ & $-(C_F - \frac{1}{2} C_A) {\cal S}$
\\
(j) & $0$ & $-(C_F - \frac{1}{2} C_A) {\cal S}$
\\
(k) -- (n$'$) & $0$ & $0$
\end{tabular}
\end{ruledtabular}
\end{table}

\subsection{Identifying the Sudakov double logarithms 
\label{sec:ident-sudakov}}

Now, let us discuss a streamlined method for identifying the Sudakov
double logarithms. The discussion in Sec.~\ref{sec:evaluation} shows
that the Sudakov double logarithms arise from the scalar integral $\cal
S$ and come from the region of integration in which a gluon, which we
label with the momentum $k$, is simultaneously soft and collinear. In
the soft approximation, we can simplify the amplitude numerators. For
example, we can write a quark-gluon vertex and the surrounding
propagator and spin-projector numerators as
\begin{equation}
{\epsilon\sl}^* (p\sl+m_c)\gamma^\mu(p\sl-k\sl+m_c)\approx 2p^\mu
{\epsilon\sl}^* (p\sl+m_c),
\label{soft-approx-num}
\end{equation}
where $\mu$ is the polarization of the gluon, we have used $p\sl^2 =
m_c^2$, and we have dropped the term that is proportional to $k$. By making
similar manipulations,  we find that, in the soft approximation, we can
always replace the quark-gluon vertex and an adjacent propagator
numerator with $\pm 2 p^\mu$ or $\pm 2 \bar p^\mu$, where the $+$ ($-$)
sign applies when the soft gluon is attached to the quark (antiquark)
line. Therefore, a Sudakov double logarithm can only arise if the gluon
with momentum $k$ attaches at one end to a collinear-to-plus-quark line
(containing momentum $p$) and at the other end to a
collinear-to-minus-quark line (containing momentum $\bar p$). Otherwise,
the contribution is subleading in $Q$ because the numerator acquires a
factor of $p^\mu p_\mu = m_c^2$ or $\bar p^\mu \bar p_\mu = m_c^2$.
Using this reasoning, we can see that diagrams (b)--(d) of 
Fig.~\ref{fig:oneloop} cannot contribute Sudakov double logarithms when the 
outer gluon is soft.

If a soft gluon enters a propagator that is off shell by order $Q^2$,
then the result is also power suppressed. For this reason, diagrams (b)--(d) 
of Fig.~\ref{fig:oneloop} cannot contribute Sudakov double logarithms 
when the inner gluon is soft, and diagrams (o)--(w) of 
Fig.~\ref{fig:oneloop_nodl} cannot contribute 
Sudakov double logarithms when the gluon that does not connect to a
spectator line is soft. For the same reason, the diagrams (x)--(y$''$) 
of Fig.~\ref{fig:oneloop_nodl} cannot contribute Sudakov double
logarithms when the vacuum-polarization gluon is soft.

Diagrams (k) and (l) of Fig.~\ref{fig:oneloop} contain only gluons that 
are connected to propagators that are off shell by order $Q^2$ and, so, 
do not contribute Sudakov double logarithms.

We conclude that only diagrams (a), (e)--(j), and (m)--(n$'$) of
Fig.~\ref{fig:oneloop} can
contribute Sudakov double logarithms, in accordance with
Table~\ref{Table:Double_Logarithms}.

\subsection{Identifying the end-point double logarithms 
\label{sec:ident-end-point}}

We can also streamline the method for identifying the end-point double
logarithms. The end-point double logarithms arise when a
spectator-quark propagator carries a momentum that is soft and
collinear. This can happen only when a gluon carries away almost all
of the spectator momentum $p$, which arises from the $J/\psi$, and a
second gluon carries away almost all of the spectator momentum $\bar p$,
which arises from the $\eta_c$. That possibility exists only for the
diagrams of Figs.~\ref{fig:oneloop}(a)--\ref{fig:oneloop}(f), 
\ref{fig:oneloop}(k), and \ref{fig:oneloop}(l). For each of 
these diagrams, we can reproduce the end-point double logarithm that is
given in Table~\ref{Table:Double_Logarithms} by using a soft
approximation for the momentum of the spectator quark $\ell$.
Specifically, we neglect $\ell^2$, $\ell \cdot p$, and $\ell \cdot \bar
p$ in comparison with $p \cdot \bar p$ in both numerators and
denominators. In principle, for the purposes of extracting end-point
logarithms, we could neglect $\ell^2$ in comparison with $\ell \cdot p$
and $\ell \cdot \bar p$ in denominators. However, we retain $\ell^2$ in
denominators in order to maintain the UV finiteness of the integrals. We
also neglect $m_{c}^2$ in comparison with $p \cdot \bar p$.

The diagram of Fig.~\ref{fig:oneloop}(a) gives
\begin{eqnarray}
i {\cal A}_{\rm NLO}^{\rm (a)\mu} &\approx& 
i {\cal A}_{\rm LO}^{\mu} 
\frac{-i \alpha_s \pi Q^2}{2} 
\left( C_F - \frac{C_A}{2} \right) 
\int_\ell 
\frac{
4 p\cdot\bar{p}(p\cdot\bar{p}-\ell\cdot \bar{p}-\ell^2)
+4(\ell\cdot p)^2-2\ell^2\,\ell\cdot p
}
{(\ell^2 -m_c^2 +i \varepsilon)
[(\ell+p)^2 +i \varepsilon]
[(\ell-\bar p)^2 +i \varepsilon]
}
\nonumber \\ 
&&
\times 
\frac{1}{
[(\ell-p-\bar p)^2 -m_c^2+i \varepsilon]
[(\ell+p+\bar p)^2 -m_c^2+i \varepsilon]},
\end{eqnarray}
where, in the numerator, we have ignored terms of order
$m_{c}^2/Q^2$ or higher. Then, the soft approximation gives, up to
corrections of order $m_c^2/Q^2$,
\begin{eqnarray}
i {\cal A}_{\rm NLO}^{\rm (a)\mu,\;soft} &=& 
i {\cal A}_{\rm LO}^{\mu} 
\frac{-i \alpha_s \pi Q^2}{2} 
\left( C_F - \frac{C_A}{2} \right)
\int_\ell 
\frac{1
}
{(\ell^2 -m_c^2 +i \varepsilon)
[(\ell+p)^2+i \varepsilon]
[(\ell-\bar p)^2+i \varepsilon]
}
\nonumber \\
&=& 
i {\cal A}_{\rm LO}^{\mu} 
\frac{-i \alpha_s \pi Q^2}{2} 
\left( C_F - \frac{C_A}{2} \right)
{\cal E}.
\end{eqnarray}

The diagram of Fig.~\ref{fig:oneloop}(b) gives
\begin{eqnarray}
i {\cal A}_{\rm NLO}^{\rm (b)\mu} &\approx& 
i {\cal A}_{\rm LO}^{\mu} 
\frac{-i \alpha_s \pi Q^2}{2}\,C_F
\nonumber \\ && 
\times 
\int_\ell 
\frac{4\ell \cdot p + 2\ell \cdot p\,\ell \cdot \bar p/p \cdot \bar p}
{(\ell^2 -m_c^2 +i \varepsilon)
[(\ell+p)^2 +i \varepsilon]
[(\ell-\bar p)^2 +i \varepsilon]
[(\ell+2 p)^2- m_c^2+i \varepsilon]
},\phantom{xxx} 
\end{eqnarray}
where we have neglected corrections of order $m_c^2/Q^2$. 
Applying the soft approximation, we obtain
\begin{eqnarray}
i {\cal A}_{\rm NLO}^{\rm (b)\mu,\;soft} &=& 
i {\cal A}_{\rm LO}^{\mu} 
\frac{-i \alpha_s \pi Q^2}{2}\, C_F
\int_\ell 
\frac{1}
{(\ell^2 -m_c^2 +i \varepsilon)
[(\ell+p)^2 +i \varepsilon]
[(\ell-\bar p)^2 +i \varepsilon]
} 
\nonumber \\ && 
\times 
\frac{4 \ell \cdot p}{\ell^2 + 4 \ell \cdot p + 3 m_c^2+i \varepsilon}.
\end{eqnarray}
Since this integral is cut off at small $\ell^\mu$ at a scale of 
order $m_c$, 
the numerator factor $4 \ell \cdot p$ cancels the last denominator 
factor, up to corrections of order $m_c^2/Q^2$. Then, we have
\begin{eqnarray}
i {\cal A}_{\rm NLO}^{\rm (b)\mu,\;soft} &\approx& 
i {\cal A}_{\rm LO}^{\mu} 
\frac{-i \alpha_s \pi Q^2}{2}\, C_F
\int_\ell 
\frac{1}
{(\ell^2 -m_c^2 +i \varepsilon)
[(\ell+p)^2 +i \varepsilon]
[(\ell-\bar p)^2 +i \varepsilon]
} 
\nonumber \\ &=& 
i {\cal A}_{\rm LO}^{\mu} 
\frac{-i \alpha_s \pi Q^2}{2}\, C_F {\cal E}.
\end{eqnarray}

The diagram of Fig.~\ref{fig:oneloop}(c) gives a contribution that is
similar to the one from the diagram of Fig.~\ref{fig:oneloop}(b):
\begin{eqnarray}
i {\cal A}_{\rm NLO}^{\rm (c)\mu,\;soft} &=& 
i {\cal A}_{\rm LO}^{\mu} 
\frac{-i \alpha_s \pi Q^2}{2} 
\,2C_F
\int_\ell 
\frac{1}
{(\ell^2 -m_c^2 +i \varepsilon)
[(\ell+p)^2 +i \varepsilon]
[(\ell-\bar p)^2 +i \varepsilon]
} 
\nonumber \\ && 
\times 
\frac{-4 \ell \cdot \bar p}
{\ell^2-4 \ell \cdot \bar p + 3 m_c^2+i \varepsilon}
\nonumber \\ &\approx& 
i {\cal A}_{\rm LO}^{\mu} 
\frac{-i \alpha_s \pi Q^2}{2} 
\,2C_F
\int_\ell 
\frac{1}
{(\ell^2 -m_c^2 +i \varepsilon)
[(\ell+p)^2 +i \varepsilon]
[(\ell-\bar p)^2 +i \varepsilon]
}
\nonumber \\
&=& 
i {\cal A}_{\rm LO}^{\mu} 
\frac{-i \alpha_s \pi Q^2}{2} 
\,2C_F {\cal E}.
\end{eqnarray}

The diagram in Fig.~\ref{fig:oneloop}(d) yields
\begin{eqnarray}
i {\cal A}_{\rm NLO}^{\rm (d)\mu,\;soft} &=& 
i {\cal A}_{\rm LO}^{\mu} 
\frac{-i \alpha_s \pi Q^2}{2} 
\,\frac{C_F}{2}
\int_\ell 
\frac{1}
{(\ell^2 -m_c^2 +i \varepsilon)
[(\ell+p)^2 +i \varepsilon]
[(\ell-\bar p)^2 +i \varepsilon]
} 
\nonumber \\ && 
\times 
\frac{8 p \cdot \bar p \; \ell^2 - 16 \ell \cdot p\, \ell \cdot \bar p}
{
(\ell^2 + 4 \ell \cdot p + 3 m_c^2+i \varepsilon)
(\ell^2-4 \ell \cdot \bar p + 3 m_c^2+i \varepsilon)
}.\label{fig2d-int}
\end{eqnarray}
In the limit $m_c\to 0$, the first term in the numerator gives the 
integral
\begin{equation}
\int_\ell \frac{1}{(\ell^2 +2 \ell \cdot p +i \varepsilon)
(\ell^2-2 \ell \cdot \bar p +i \varepsilon)
(\ell^2+4 \ell \cdot p +i \varepsilon)
(\ell^2-4 \ell \cdot \bar p +i \varepsilon)}. 
\end{equation}
This integral has a logarithmically divergent soft power count and a
logarithmically divergent collinear power count. However, in the soft
region, in which we can neglect $\ell^2$ in comparison with $\ell\cdot
p$ or $\ell\cdot \bar p$, there is no pinch in either the $\ell^+$ or
$\ell^-$ contour of integration. Therefore this integral does not give
an end-point double logarithm. Retaining the second term in the numerator
in Eq.~(\ref{fig2d-int}), we obtain, up to corrections of order
$m_c^2/Q^2$,
\begin{eqnarray}
i {\cal A}_{\rm NLO}^{\rm (d)\mu,\;soft} &\approx& 
i {\cal A}_{\rm LO}^{\mu} 
\frac{-i \alpha_s \pi Q^2}{2} 
\,\frac{C_F}{2}
\int_\ell 
\frac{1}
{(\ell^2 -m_c^2 +i \varepsilon)
[(\ell+p)^2 +i \varepsilon]
[(\ell-\bar p)^2 +i \varepsilon]
}
\nonumber \\
&=& 
i {\cal A}_{\rm LO}^\mu 
\frac{-i \alpha_s \pi Q^2}{2} 
\,\frac{C_F}{2} {\cal E}. 
\end{eqnarray}

The contribution of the diagram in Fig.~\ref{fig:oneloop}(e) can be
evaluated in a similar manner:
\begin{eqnarray}
i {\cal A}_{\rm NLO}^{\rm (e)\mu,\;soft} &\approx& 
i {\cal A}_{\rm LO}^{\mu} 
\frac{-i \alpha_s \pi Q^2}{2} 
\left( C_F - \frac{C_A}{2} \right) 
\int_\ell 
\frac{1}
{(\ell^2 -m_c^2 +i \varepsilon)
[(\ell+p)^2 +i \varepsilon]
[(\ell-\bar p)^2 +i \varepsilon]
}
\nonumber \\
&=& 
i {\cal A}_{\rm LO}^{\mu} 
\frac{-i \alpha_s \pi Q^2}{2} 
\left( C_F - \frac{C_A}{2}  \right) 
{\cal E}. 
\end{eqnarray}

The contribution of the diagram in Fig.~\ref{fig:oneloop}(f) is, 
up to corrections of order $m_c^2/Q^2$, 
\begin{eqnarray}
i {\cal A}_{\rm NLO}^{\rm (f)\mu,\;soft} &=& 
i {\cal A}_{\rm LO}^{\mu} 
\frac{-i \alpha_s \pi Q^2}{2} 
\left( C_F - \frac{C_A}{2}  \right) 
\int_\ell 
\frac{1}
{(\ell^2 -m_c^2 +i \varepsilon)
[(\ell+p)^2 +i \varepsilon]
[(\ell-\bar p)^2 +i \varepsilon]
} 
\nonumber \\ &&
\times 
\frac{2\ell\cdot p-2\ell\cdot \bar p}
{2p\cdot \bar p}.
\end{eqnarray}
Since all of the numerator terms are proportional to $\ell$, this 
integral does not give a divergent soft 
power count in the limit $m_c\to 0$. Hence, it does not contribute an 
end-point double logarithm.

In the same manner, the diagrams in Figs.~\ref{fig:oneloop}(k) and 
\ref{fig:oneloop}(l)
lead to integrals that do not give divergent soft power counts and,
therefore, do not contribute end-point double logarithms.

\section{General analysis of the Sudakov double 
logarithms\label{sec:sudakov}}

As we have mentioned, the singularities that arise from the Sudakov
double logarithms of $Q^2/m_c^2$ as $m_c\to 0$ come from a region of loop
momentum in which the momentum of a gluon is simultaneously soft and collinear.
Consequently, we can organize these singularities by making use of
soft and/or collinear approximations for the amplitudes.

Consider, for example, Figs.~\ref{fig:oneloop}(a) and \ref{fig:oneloop}(g), 
for the situation in which the gluon with momentum $k$ is collinear to plus.
Then, the upper quark-gluon vertex and the propagator and spin-projector
numerator factors surrounding it can be written as
\begin{equation}
(-p\sl-k\sl +m_c)
\gamma^\nu
( -p\sl+m_c) {\epsilon\sl}^*
\approx -2(p + k)^\nu(-p\sl +m_c) {\epsilon\sl}^* +m_c \gamma^\nu k\sl
{\epsilon\sl}^*,
\label{collinear-to-plus}
\end{equation}
where $\nu$ is the polarization index of the gluon, we have used $k\sl
p\sl \propto p\sl^2=m_c^2$, and we have dropped terms of order $m_c^2$.
In the case of massless quarks, one obtains the collinear-to-plus
approximation by retaining only the first term on the right side of
Eq.~(\ref{collinear-to-plus}). Since that term is proportional to
$k^\nu$ for $k$ collinear to $p$, one can use graphical Ward identities
to simplify the amplitude. In the case of nonzero quark masses, the
second term in Eq.~(\ref{collinear-to-plus}) generally spoils this
approach. However, if $k$ is soft in comparison with $p$, as well as
collinear, then we can drop the second term on the right side of
Eq.~(\ref{collinear-to-plus}), and the standard collinear-to-plus
approximation holds. Since the current in Eq.~(\ref{collinear-to-plus})
now lies in the plus light-cone direction, up to terms of order $m_c^2$,
we can make a collinear-to-plus approximation in the gluon
propagator~\cite{Bodwin:1984hc, Collins:1985ue, Collins:1989gx} by
making the following replacement in the gluon polarization tensor:
\begin{equation}
g_{\mu\nu}\to \frac{k_\mu \bar n_\nu}{k\cdot \bar n- i\varepsilon},
\label{coll-prop-plus}
\end{equation}
where $\bar n$ is a unit vector in the minus light-cone direction, and the
index $\nu$ corresponds to the upper attachment of the gluon to the
quark line with momentum $-p$. The sign of $i\varepsilon$ is fixed by
the sign in the $\mu$-side fermion propagator in the original Feynman
diagram. (We always choose $k$ to flow out of collinear-to-plus lines
and into collinear-to-minus lines.) This soft-collinear-to-plus
approximation is valid unless the $\mu$ attachment of the gluon is to a
line that is also collinear to plus. Hence, the approximation always
holds, in the soft-collinear limit, for the diagrams that produce
Sudakov logarithms because the invariant $Q^2$ in the logarithm can
appear only if the soft-collinear gluon connects a line carrying
momentum $p$ to a line carrying momentum $\bar p$.

In a similar fashion, if $k$ is collinear to minus, then we can make the 
replacement
\begin{equation}
g_{\mu\nu}\to \frac{k_\mu n_\nu}{k\cdot n+ i\varepsilon},
\label{coll-prop-minus}
\end{equation}
where $n$ is a unit vector in the plus light-cone direction.

The replacement (\ref{coll-prop-plus}) can also be regarded as a soft
approximation~\cite{Grammer:1973db, Collins:1981uk} to the $\mu$
attachment of the gluon. For example, in the case of the diagram of
Fig.~\ref{fig:oneloop}(a), the lower vertex of the gluon with momentum
$k$ and the surrounding propagator and spin-projector factors can be
written in the soft approximation as
\begin{equation}
(-\bar p\sl+k\sl+m_c)\gamma^\mu(-\bar p\sl+m_c)\gamma^5 \approx -2\bar p^\mu 
(-\bar p\sl+m_c)\gamma^5,
\end{equation}
where we have neglected $k$ relative to $\bar p$. Then, the replacement
(\ref{coll-prop-plus}) is also valid by virtue of the fact that, in  
the limit $m_c\to 0$, the current
at lower vertex of the gluon lies in the minus light-cone direction. 

The soft approximation applies only when the current on the $\mu$ side
of the gluon is in the minus light-cone direction. In contrast, the
soft-collinear-to-plus approximation applies whenever the current on the
$\mu$ side of the gluon is different from the plus direction, and so the
collinear approximation can be more versatile than the soft
approximation.

For the diagram of Fig.~\ref{fig:oneloop}(a) we can apply the
soft-collinear-to-plus approximation (\ref{coll-prop-plus}) to the lower vertex
of the gluon with momentum $k$ and then make use of the graphical Ward 
identity (the Feynman identity)
\begin{equation}
k\sl =(-\bar p\sl+k\sl-m_c)-(-\bar p\sl -m_c).
\end{equation}
The result, including the adjacent propagator and the factor $(-\bar
p\sl+m_c)\gamma^5$ from the spin projector, is
\begin{equation}
\frac{1}{-\bar p\sl+k\sl-m_c+i\varepsilon}\frac{k\sl \bar n_\nu}{k\cdot \bar
n-i\varepsilon} (-\bar p\sl+m_c)\gamma^5=\frac{
\bar n_\nu}{k\cdot \bar n-i\varepsilon} 
(-\bar p\sl+m_c)\gamma^5.
\end{equation}
Similarly, for the diagram of Fig.~\ref{fig:oneloop}(g), we can
apply the soft-collinear-to-plus approximation (\ref{coll-prop-plus}) to the lower
vertex of the gluon with momentum $k$ and make use of the Feynman
identity to obtain
\begin{equation}
\gamma^5 (\bar p\sl+m_c) \frac{k\sl\bar n_\nu}{k\cdot \bar n-i\varepsilon}
\frac{1}{\bar p\sl-k\sl-m_c+i\varepsilon}
= 
- \frac{\bar n_\nu}{k\cdot \bar n
-i\varepsilon}
\gamma^5 (\bar p\sl+m_c).
\end{equation}

Now, in the soft approximation, we can neglect $k$ in comparison with
$p$ and $\bar p$ in quark propagator numerators, and we can neglect all
of the invariants involving $k$ in comparison with $p \cdot \bar p$ in
propagator denominators. Therefore, in the diagram of
Fig.~\ref{fig:oneloop}(a), we can neglect $k$ in the gluon propagator
that has momentum $p+\bar p +k$ and in the quark propagator that has
momentum $2p+\bar p+k$. Then, in the soft-collinear-to-plus
approximation, the contributions of the diagrams of
Figs.~\ref{fig:oneloop}(a) and \ref{fig:oneloop}(g) have exactly the same 
propagator and
vertex factors, except for a relative minus sign. Furthermore, because
the outgoing $c\bar c$ pairs are in color-singlet states, the
diagrams of Figs.~\ref{fig:oneloop}(a) and \ref{fig:oneloop}(g) have the same 
color factors. Therefore, the contributions of these diagrams cancel when the
gluon with momentum $k$ is soft and collinear to plus.

This type of pairwise cancellation occurs for all of the
soft-collinear-to-plus contributions from diagrams 
(a) and (e)--(j) of Fig.~\ref{fig:oneloop}, including
contributions from the diagram of Fig.~\ref{fig:oneloop}(a), in which the
gluon with momentum $p+\bar p+k$ is soft and collinear to plus. Similar
cancellations occur for the diagrams 
(a) and (e)--(j) of Fig.~\ref{fig:oneloop}  
when a gluon has soft-collinear-to-minus momentum.

Now consider the diagrams of Figs.~\ref{fig:oneloop}(m) and \ref{fig:oneloop}(m$'$). In
each diagram, we apply the soft-collinear-to-plus approximation to the
lower vertex of the gluon whose upper vertex connects to the
spectator-quark. The resulting expressions for the two diagrams
differ only by a minus sign and a color factor. Let $T^a$ be the color
matrix that is associated with the lower vertex of the gluon to which
soft-collinear-to-plus approximation is applied and let $T^b$ be the
color matrix that is associated with the lower vertex of the other
gluon. Then, the diagram of Fig.~\ref{fig:oneloop}(m) contributes
$T^aT^b$ and the diagram of Fig.~\ref{fig:oneloop}(m$'$) contributes
$-T^bT^a$ to the color factor on the lower quark line. These
contributions combine to give $[T^a, T^b]=if^{abc}T^c$, which has zero
overlap with the color-singlet lower meson. [Here, $f^{abc}$ is the
structure constant of $SU(3)$.] In each diagram, we can also
apply the soft-collinear-to-minus approximation to the upper vertex of the
gluon whose lower vertex connects to the spectator-quark line.
Again, we obtain a vanishing contribution. Similar arguments show
that the soft-collinear-to-plus and soft-collinear-to-minus 
contributions from the diagrams of Figs.~\ref{fig:oneloop}(n) and \ref{fig:oneloop}(n$'$)
cancel.

These arguments go through in the same fashion for the charge-conjugate
diagrams. We conclude that, in the sum of all diagrams, Sudakov double
logarithms cancel, verifying the results of our explicit calculation. 

\begin{figure}
\epsfig{file=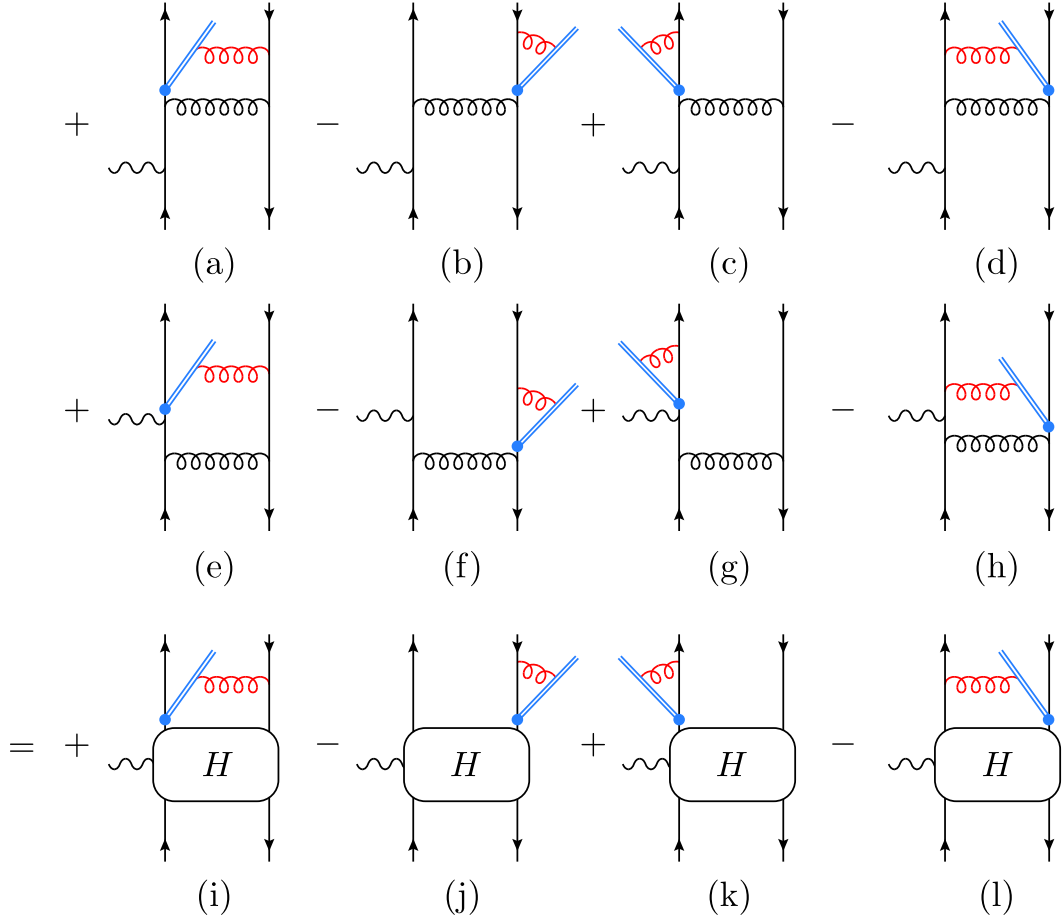,width=14cm}
\caption{Contributions that arise from regions of integration in which 
a gluon momentum is soft and collinear. The double lines are 
eikonal lines, whose Feynman rules are described in the text. $H$ 
denotes the hard subdiagram. We have 
not shown diagrams that can be obtained by charge conjugation or 
diagrams that can be obtained by interchanging the upper and lower 
mesons. \label{fig:dist-functions}
}
\end{figure}

We note that, after the application of the soft-collinear-to-plus
approximation and the Feynman identity, the soft-collinear-to-plus
contributions take the forms that are shown in
Fig.~\ref{fig:dist-functions}. Here, the double line is an eikonal
line, with gluon-eikonal vertex $\bar n_\nu$ and eikonal propagator
$-i/(k\cdot \bar n-i\varepsilon)$, which arise from the replacement
(\ref{coll-prop-plus}). (There is no propagator between the attachment
of the eikonal line to the quark line and the adjacent vertex on the
quark line.) For example, the contribution of
Fig.~\ref{fig:dist-functions}(a) arises from the contribution of
Fig.~\ref{fig:oneloop}(a) in which the gluon with momentum $k$ is soft
and collinear to plus, and the contribution of
Fig.~\ref{fig:dist-functions}(b) arises from the contribution of
Fig.~\ref{fig:oneloop}(g) in which the gluon with momentum $k$ is
soft and collinear to plus. Now, the pairwise cancellation of these
contributions is obvious from the diagrams in
Fig.~\ref{fig:dist-functions}. All of the soft-collinear-to-plus (-minus)
contributions are associated entirely with the upper (lower) meson and
have the form of contributions to the meson distribution amplitude. We
define the hard subdiagram to be the subdiagram in which all propagators
are off shell by order $Q^2$. Then, as is depicted in
Figs.~\ref{fig:dist-functions}(i)--\ref{fig:dist-functions}(l), the soft-collinear contributions
have been factored from the hard subdiagram.

In arriving at the forms in  Fig.~\ref{fig:dist-functions}, we have
made use of the fact that a soft momentum can be neglected in the hard
subdiagram and the fact that both of the mesons are color-singlet
states. We can arrive at the forms in  Fig.~\ref{fig:dist-functions}
without making these assumptions, as follows. As we have already
mentioned, the replacement (\ref{coll-prop-plus}) applies whenever the
$\mu$ vertex of the soft-collinear-to-plus gluon attaches to a line that
is not collinear to plus. Therefore, we can apply the replacement
(\ref{coll-prop-plus}) to diagrams in which the soft-collinear gluon
enters the hard subdiagram. These diagrams give a vanishing contribution
in the soft limit. Nevertheless, we can include them formally. Then, we
can apply the diagrammatic Ward identities to arrive at the forms in
Fig.~\ref{fig:dist-functions} directly. For example, the contributions
from diagrams of Figs.~\ref{fig:oneloop}(a), \ref{fig:oneloop}(e), 
and \ref{fig:oneloop}(l) in which the
gluon with momentum $k$ is soft and collinear to plus combine to give the
contribution of the diagram in Fig.~\ref{fig:dist-functions}(a).
Discussions of the required non-Abelian diagrammatic Ward identities can
be found, for example, in Refs.~\cite{Bodwin:1984hc,Collins:1989gx}.

The factorized forms in Figs.~\ref{fig:dist-functions}(i)--\ref{fig:dist-functions}(l) rely on
the soft-collinear approximation and the diagrammatic Ward identities.
The cancellations of the Sudakov logarithms then follow from the
color-singlet natures of the mesons. Given the general natures of these
arguments, we expect them to hold to all orders in perturbation theory.
For the case of massless quarks, all-orders arguments for the
cancellation of Sudakov logarithms in exclusive meson amplitudes have
been given in
Refs.~\cite{Cornwall:1975ty,Efremov:1978rn,Duncan:1979hi,Duncan:1979ny,Efremov:1979qk}.
The all-orders cancellation of soft divergences in exclusive meson
amplitudes can also be established for the case of massless quarks in the
context of soft-collinear effective theory (SCET) by making use of a
field redefinition \cite{Bauer:2001yt}.

\section{General analysis of the end-point region \label{sec:end-point}}

As we have mentioned, the singularities that arise 
from the end-point double logarithms of $Q^2/m_{c}^2$ as $m_c \to 0$ come 
from the region of
loop integration in which the momentum $\ell$ of the internal
spectator-quark line is simultaneously soft and collinear. Hence, in
order to analyze those singularities, we need to consider only diagrams
that can give rise to such a momentum configuration. A necessary
condition is that both gluons attach to the spectator line. The diagrams
satisfying this condition are shown in
Figs.~\ref{fig:oneloop}(a)--\ref{fig:oneloop}(f), \ref{fig:oneloop}(k),
and \ref{fig:oneloop}(l).

At leading order in $m_c/Q$, power counting arguments
\cite{Collins:1989gx} show that there are no logarithmic singularities
as $m_c\to 0$ that arise from the region $\ell\to 0$. Therefore, the
end-point double logarithms appear only in contributions in which there
is at least one numerator factor $m_c$, {\it i.e.}, 
a helicity flip. 

We now argue that, if a contribution is to produce a double logarithm at the
first subleading order in $m_c/Q$,
then it must contain exactly one numerator factor $m_c$. First, we note
that the contribution must contain an odd number of numerator factors of
$m_c$ in order to produce the helicity flip that is required by the
process $e^+e^-\to J/\psi +\eta_c$. If there are three or more numerator
factors of $m_c$, then either the contribution is suppressed by powers
of $m_c/Q$ or the integration produces two or more inverse powers of
$m_c$. In the latter case, the inverse powers of $m_c$ must arise from
either the soft singularity or the collinear singularity. The
singularity that produces the inverse power of $m_c$ cannot produce a
logarithm. Therefore, such contributions contain, at most, a single
logarithm of $m_c$.\footnote{Detailed power counting arguments show that
the soft singularity can be, at most, quadratically divergent, while the
collinear singularities can be, at most, linearly divergent. This
situation occurs only in the case of the diagram of
Fig.~\ref{fig:oneloop}(d). Hence, a single logarithm at the first subleading
order in $m_c/Q$ can arise only if
the soft singularity produces two inverse powers of $m_c$ and the
collinear singularity produces a logarithm.}  In principle,
contributions that contain a single numerator factor of $m_c$ can
diverge as inverse powers of $m_c$ in the limit $m_c\to 0$, but, as we
will see, such contributions vanish when the numerator trace is taken.

In the diagrams of Figs.~\ref{fig:oneloop}(a), \ref{fig:oneloop}(e), \ref{fig:oneloop}(f), \ref{fig:oneloop}(k), and \ref{fig:oneloop}(l),
the momenta of the propagators on the active-quark lines contain both $p$
and $\bar p$. Since $p\cdot \bar p\sim P^2\sim Q^2$, we can ignore
$\ell$ in the denominators of those propagators. In the limit $m_c\to
0$, the two gluon-propagator denominators and the
spectator-quark-propagator denominator produce factors
$1/(\ell^2+2p\cdot \ell+i \varepsilon)$, $1/(\ell^2-2\bar p\cdot \ell+i
\varepsilon)$, and $1/(\ell^2+i \varepsilon)$, respectively, where we
have dropped the $m_c^2$ terms in the propagator denominators. Hence, if
we are to obtain a logarithmically divergent soft power count
($\lambda^{-4}$), then we cannot have any numerator factors of $\ell$.
This implies that we must retain the factor $m_c$ in the numerator of
the spectator-quark propagator and that we can drop $m_c$ elsewhere in
the numerator.

In the diagram of Fig.~\ref{fig:oneloop}(b), the denominator of the
outermost active-quark propagator produces a factor $1/(\ell^2+4p\cdot
\ell+i \varepsilon)$ in the limit $m_c\to 0$. Hence, taking into account
the two gluon-propagator denominators and the spectator-quark-propagator
denominator, we see that the propagator denominators, by themselves,
produce a linearly divergent soft power count and a linearly divergent
collinear-to-plus power count. However, as we now show, the numerator
factors reduce both of these power counts to logarithmic ones. First, we
rewrite the numerator factors that are associated with the outermost
gluon and the spin projector for the $J/\psi$ as
\begin{equation}
\gamma_\mu (p\sl
-m_c){\epsilon\sl}^*\gamma^\mu=2m_c{\epsilon\sl}^*,
\label{projector-gluon}
\end{equation}
where we have used the fact that $p\cdot \epsilon^*=0$. Now, because
this factor contains the numerator power of $m_c$, the numerator of the
spectator-quark propagator must contribute a factor $\ell\sl$.
Furthermore, if $\ell$ is proportional to $p$, then the upper active
or spectator propagator combines with the expression
(\ref{projector-gluon}) to give the structure $p\sl \, \epsilon\sl^*
p\sl=-\epsilon\sl^* p^2=-\epsilon\sl^*m_c^2$. That is, the numerator
vanishes, up to terms of order $m_c^2$. This implies that, in the trace
over the gamma matrices, $\ell$ must appear in the combination
$\ell\cdot p$. This numerator factor reduces both the soft and the collinear
power counts to logarithmic ones.

In the case of the diagram of Fig.~\ref{fig:oneloop}(c), the
denominator of the outermost active-quark propagator contributes a
factor $1/(\ell^2-4\bar p\cdot \ell+i \varepsilon)$. Hence, taking into
account the two gluon-propagator denominators and the
spectator-quark-propagator denominator, we see that the propagator
denominators, by themselves, produce linearly divergent soft and
collinear-to-minus power counts. We rewrite the numerator factors that
are associated with the outermost gluon and the spin projector for the
$\eta_c$ as
\begin{equation}
\gamma_\mu (-{\bar p}\sl-m_c)\gamma^5\gamma^\mu =(-2{\bar p}\sl
+4m_c)\gamma^5.
\label{projector-gluon2}
\end{equation}
Now, it is easy to see that there must be a numerator factor of $\ell$
from either the outermost active-quark propagator or the spectator-quark
propagator. Otherwise, there will be two factors of ${\bar p}\sl$ that
are either adjacent or separated by $\gamma^5$, resulting in an
expression that vanishes, up to terms of order $m_c^2$. Furthermore, if
$\ell$ is proportional to $\bar p$, then the numerator vanishes, up to
terms of order $m_c^2$, because we again have a situation in which two
factors of ${\bar p}\sl$ are either adjacent or separated by $\gamma^5$.
Therefore, $\ell$ must appear in the combination $\ell\cdot \bar p$ in
the trace over gamma matrices. The factor $\ell\cdot \bar p$ reduces
both the soft and collinear-to-minus power counts to logarithmic ones.

In the case of the diagram of Fig.~\ref{fig:oneloop}(d), the denominators
of the active-quark propagators produce factors $1/(\ell^2+4p\cdot
\ell+i \varepsilon)$ and $1/(\ell^2-4\bar p\cdot \ell+i \varepsilon)$ in
the limit $m_c\to 0$. Hence, taking into account the two
gluon-propagator denominators and the spectator-quark-propagator
denominator, we see that the propagator denominators, by themselves,
produce a quadratically divergent soft power count and linearly
divergent collinear-to-plus and collinear-to-minus power counts. One can
apply the arguments that were used for the numerators of the diagrams of
Figs.~\ref{fig:oneloop}(b) and \ref{fig:oneloop}(c) separately to each of the 
gluons
in the diagram of Fig.~\ref{fig:oneloop}(d). The conclusion is that the
numerator contains two factors of $\ell\sl$ and that the numerator
vanishes, up to terms of order $m_c^2$, if $\ell$ is proportional to $p$
or to $\bar p$. Therefore, the trace contains a factor $\ell\cdot p\,
\ell\cdot \bar p$ or a factor $\ell^2$. Either factor reduces the soft
and collinear power counts to logarithmic ones. (In fact, there is no 
collinear pinch if the numerator factor is $\ell^2$.)

We note that the appearance of singularities that arise from the
regions in which $\ell$ is soft or soft collinear relies on the presence
of a numerator factor $m_c$. Consider, for example, the process in which
a virtual photon produces a spin-zero, $S$-wave meson ($\eta_c$) and a
spin-zero, $P$-wave meson ($h_c$). This process proceeds without a
helicity flip. Therefore, at LO in $m_c/Q$, we can set $m_c=0$
everywhere in the numerator. Then, for each of the contributions of the
diagrams of Figs.~\ref{fig:oneloop}(a)--\ref{fig:oneloop}(f), \ref{fig:oneloop}(k), and \ref{fig:oneloop}(l), there must be
a numerator factor of $\ell$ from the spectator-quark propagator. In
addition, by making use of the identity (\ref{projector-gluon2}) with
$m_c=0$, we can see that the contributions of the diagrams of
Figs.~\ref{fig:oneloop}(b) and \ref{fig:oneloop}(c) contain an additional factor of
$\ell$ from the outermost active-quark propagator and that the
contribution of the diagram of Fig.~\ref{fig:oneloop}(d) contains two
additional factors of $\ell$ from the two active-quark propagators.
Otherwise, these contributions would vanish because there would be two
factors of $p\sl$ or two factors of $\bar p\sl$ that are adjacent or are
separated by $\gamma^5$. These numerator factors of $\ell$ are
sufficient to eliminate the soft singularity in the contributions of
each of the diagrams of Figs.~\ref{fig:oneloop}(a)--\ref{fig:oneloop}(f), \ref{fig:oneloop}(k), and \ref{fig:oneloop}(l).
We have verified this analysis by carrying out explicit calculations of
the contributions of each diagram. Arguments that are similar to the
preceding one apply to situations in which one or both mesons are
spin-one states. The conclusion is that, at leading order in $m_c/Q$,
for processes that do not involve a helicity flip, there are no
singularities that arise from the region in which the spectator quark
carries a soft or a soft-collinear momentum (end-point singularities).
This conclusion is in agreement with explicit calculations for helicity
nonflip processes \cite{Jia:2010fw,Dong:2011fb,Jia:2008ep} and with
general analyses of leading pinch singularities \cite{Collins:1989gx}.

\section{Summary}
\label{sec:summary}%

In this paper we have analyzed double logarithms of $Q^2/m_c^2$ that
appear in the NLO QCD corrections to the process $e^+ e^- \to
J/\psi+\eta_c$. We have identified the origins of these double
logarithms by examining, in the limit $m_c\to 0$, the pinch
singularities in the contours of integration in the amplitudes.  We
have found that the double logarithms are of two types: Sudakov double
logarithms and end-point double logarithms. The Sudakov double logarithms
are characterized by singularities in the limit $m_c\to 0$ that arise
from a momentum region in which the momentum of a gluon is both soft
and collinear to one of the outgoing mesons. The end-point double
logarithms are characterized by singularities in the limit $m_c\to 0$
that arise from a momentum region in which one gluon carries away
almost all of the momentum of a spectator quark and the other gluon
carries away almost all of the momentum of a spectator antiquark. We
have carried out an explicit calculation that shows that the Sudakov and
end-point double logarithms account for all of the double logarithms that
arise from each Feynman diagram.

When one sums over the contributions of all of the diagrams, the Sudakov
double logarithms cancel. We have shown that this cancellation can be
understood by means of a general argument that is based on a
soft-collinear approximation and graphical Ward identities.

We have found that the end-point singular region can be
interpreted as a pinch-singular region in which the momentum of the
spectator-quark line is both soft and collinear. Such a pinch-singular
region is allowed by a general Landau analysis. However, in the case of
processes that proceed without a helicity flip, it does not give rise to
a singular power count at leading order in $m_c/Q$. In the case of
processes that proceed only through a helicity flip, such as $e^+e^-\to
J/\psi + \eta_c$, the end-point singular region can give rise to a
singular power, owing to the presence of one or more factors of $m_c$
in numerators of Feynman amplitudes at the leading nontrivial order
in $m_c/Q$. We have given a general analysis of the power counting in
the end-point singular region at NLO in $\alpha_s$. That analysis
shows that the end-point singularities can produce logarithms of $m_c$, 
but not inverse powers of $m_c$. It is apparent that the end-point region can
give rise to single logarithms of $Q^2/m_c^2$, as well as double
logarithms. The single logarithms correspond to a region of
integration in which the momentum of a spectator-quark line is soft, but
not collinear. 

Finally, we remark that the insight that the end-point singular region is
a pinch-singular region in which a spectator-quark line is either soft
or soft collinear might allow one to make progress in organizing end-point
singularities to all orders in perturbation theory. We note that SCET,
as it is presently formulated \cite{Bauer:2000yr}, does not include
modes in which quarks are soft, and so it would be necessary to augment
SCET in order to apply it to the end-point region. The all-orders
organization of end-point singularities, in the context of SCET or in the
context of traditional diagrammatic approaches, would be a key
ingredient in the resummation of the end-point logarithms.

\begin{acknowledgments}

We are grateful to Yu Jia and Xiu-Ting Yang for providing us with
diagram-by-diagram results for the logarithms in the calculation of
Ref.~\cite{Jia:2010fw}. We also thank George Sterman and Jianwei Qiu for
helpful discussions on the collinear approximation in the case of
nonzero fermion mass. J.~L.\ thanks the theory group of the High Energy
Physics Division at Argonne National Laboratory for its hospitality
during his visit. The work of G.~T.~B.\ and H.~S.~C.\ is supported
by the U.S.\ Department of Energy, Division of High Energy Physics,
under contract No. DE-AC02-06CH11357. This work was supported in part by
Korea University and by APCTP through KQWG. 
The submitted manuscript has been created in part by
UChicago Argonne, LLC, operator of Argonne National Laboratory. Argonne,
a U.S.\ Department of Energy Office of Science laboratory, is operated
under contract No. DE-AC02-06CH11357. The U.S. Government retains for
itself, and others acting on its behalf, a paid-up nonexclusive,
irrevocable worldwide license in said article to reproduce, prepare
derivative works, distribute copies to the public, and perform publicly
and display publicly, by or on behalf of the Government.

\end{acknowledgments}



\begin{thebibliography}{}
\bibitem{Abe:2002rb}
  K.~Abe {\it et al.}  [Belle Collaboration],
  Phys.\ Rev.\ Lett.\  {\bf 89} (2002) 142001
  [hep-ex/0205104].

\bibitem{Abe:2004ww}
  K.~Abe {\it et al.}  [Belle Collaboration],
  Phys.\ Rev.\ D {\bf 70} (2004) 071102
  [hep-ex/0407009].

\bibitem{Aubert:2005tj}
  B.~Aubert {\it et al.}  [BABAR Collaboration],
  Phys.\ Rev.\ D {\bf 72} (2005) 031101
  [hep-ex/0506062].

\bibitem{Bodwin:1994jh}
  G.T.~Bodwin, E.Braaten, and G.P.~Lepage,
  Phys.\ Rev.\ D {\bf 51} (1995) 1125;
 {\bf 55} (1997) 5853(E)
 [hep-ph/9407339].

\bibitem{Braaten:2002fi}
  E.~Braaten and J.~Lee,
  Phys.\ Rev.\ D {\bf 67} (2003) 054007;
  {\bf 72} (2005) 099901(E)
  [hep-ph/0211085].

\bibitem{Liu:2002wq}
  K.-Y.~Liu, Z.-G.~He, and K.-T.~Chao,
  Phys.\ Lett.\ B {\bf 557} (2003) 45
  [hep-ph/0211181].

\bibitem{Zhang:2005cha}
  Y.-J.~Zhang, Y.-j.~Gao, and K.-T.~Chao,
  Phys.\ Rev.\ Lett.\  {\bf 96} (2006) 092001
  [hep-ph/0506076].

\bibitem{Gong:2007db}
  B.~Gong and J.-X.~Wang,
  Phys.\ Rev.\ D {\bf 77} (2008) 054028
  [arXiv:0712.4220 [hep-ph]].

\bibitem{Bodwin:2006dn}
  G.T.~Bodwin, D.~Kang, and J.~Lee,
  Phys.\ Rev.\ D {\bf 74} (2006) 014014
  [hep-ph/0603186].

\bibitem{Bodwin:2006ke}
  G.T.~Bodwin, D.~Kang, T.~Kim, J.~Lee, and C.~Yu,
  AIP Conf.\ Proc.\  {\bf 892} (2007) 315
  [hep-ph/0611002].

\bibitem{Bodwin:2007ga}
  G.T.~Bodwin, J.~Lee, and C.~Yu,
  Phys.\ Rev.\ D {\bf 77} (2008) 094018
  [arXiv:0710.0995 [hep-ph]].

\bibitem{Jia:2010fw}
  Y.~Jia, J.-X.~Wang, and D.~Yang,
  JHEP {\bf 1110} (2011) 105
  [arXiv:1012.6007 [hep-ph]].

\bibitem{Lepage:1980fj} 
  G.P.~Lepage and S.~J.~Brodsky,
  Phys.\ Rev.\ D {\bf 22}, 2157 (1980).

\bibitem{Chernyak:1983ej} 
  V.L.~Chernyak and A.~R.~Zhitnitsky,
  Phys.\ Rept.\  {\bf 112}, 173 (1984).

\bibitem{Bodwin:2013ys} 
  G.T.~Bodwin, H.S.~Chung, and J.~Lee,
  PoS ConfinementX {\bf }, 133 (2012)
  [arXiv:1301.3937 [hep-ph]].

\bibitem{Bodwin:2008nf}
  G.T.~Bodwin, X.~Garcia i Tormo, and J.~Lee,
  Phys.\ Rev.\ Lett.\  {\bf 101} (2008) 102002
  [arXiv:0805.3876 [hep-ph]].

\bibitem{Bodwin:2010fi}
  G.T.~Bodwin, X.~Garcia i Tormo, and J.~Lee,
  Phys.\ Rev.\ D {\bf 81} (2010) 114014
  [arXiv:1003.0061 [hep-ph]].

\bibitem{Dong:2012xx} 
  H.-R.~Dong, F.Feng, and Y.~Jia,
  Phys.\ Rev.\ D {\bf 85}, 114018 (2012)
  [arXiv:1204.4128 [hep-ph]].

\bibitem{Landau:1959fi} 
  L.D.~Landau,
  Nucl.\ Phys.\  {\bf 13}, 181 (1959).

\bibitem{Coleman:1965xm} 
  S.~Coleman and R.E.~Norton,
  Nuovo Cim.\  {\bf 38}, 438 (1965).

\bibitem{Sterman:1978bi} 
  G.F.~Sterman,
  Phys.\ Rev.\ D {\bf 17}, 2773 (1978).

\bibitem{Sterman:1978bj} 
  G.F.~Sterman,
  Phys.\ Rev.\ D {\bf 17}, 2789 (1978).

\bibitem{Libby:1978bx} 
  S.B.~Libby and G.F.~Sterman,
  Phys.\ Rev.\ D {\bf 18}, 4737 (1978).

\bibitem{Drell:1969km} 
  S.~D.~Drell and T.~-M.~Yan,
  Phys.\ Rev.\ Lett.\  {\bf 24}, 181 (1970).

\bibitem{West:1970av}
  G.~B.~West,
  Phys.\ Rev.\ Lett.\  {\bf 24}, 1206 (1970).

\bibitem{Bodwin:2002hg}
  G.T.~Bodwin and A.~Petrelli,
  Phys.\ Rev.\ D {\bf 66} (2002) 094011
  [hep-ph/0205210].

\bibitem{Sterman:2004pd} 
  See G.F.~Sterman,
  hep-ph/0412013 for a detailed discussion.

\bibitem{jia-private-comm}
Y.~Jia and X.-T.~Wang, private communication.

\bibitem{Bodwin:1984hc}
  G.T.~Bodwin,
  Phys.\ Rev.\ D {\bf 31} (1985) 2616;
  {\bf 34} (1986) 3932(E).

\bibitem{Collins:1985ue}
  J.C.~Collins, D.E.~Soper, and G.F.~Sterman,
  Nucl.\ Phys.\ B {\bf 261} (1985) 104.

\bibitem{Collins:1989gx}
  J.C.~Collins, D.E.~Soper, and G.F.~Sterman,
  Adv.\ Ser.\ Direct.\ High Energy Phys.\  {\bf 5} (1988) 1
  [hep-ph/0409313].

\bibitem{Grammer:1973db}
  G.~Grammer, Jr. and D.R.~Yennie,
  Phys.\ Rev.\ D {\bf 8} (1973) 4332.

\bibitem{Collins:1981uk}
  J.C.~Collins and D.E.~Soper,
  Nucl.\ Phys.\ B {\bf 193} (1981) 381;
  {\bf 213} (1983) 545(E).

\bibitem{Cornwall:1975ty} 
  J.~M.~Cornwall and G.~Tiktopoulos,
  Phys.\ Rev.\ D {\bf 13}, 3370 (1976).

\bibitem{Efremov:1978rn} 
  A.~V.~Efremov and A.~V.~Radyushkin,
  Theor.\ Math.\ Phys.\  {\bf 42}, 97 (1980)
  [Teor.\ Mat.\ Fiz.\  {\bf 42}, 147 (1980)].

\bibitem{Duncan:1979hi} 
  A.~Duncan and A.~H.~Mueller,
  Phys.\ Rev.\ D {\bf 21}, 1636 (1980).

\bibitem{Duncan:1979ny} 
  A.~Duncan and A.~H.~Mueller,
  Phys.\ Lett.\ B {\bf 90}, 159 (1980).

\bibitem{Efremov:1979qk} 
  A.~V.~Efremov and A.~V.~Radyushkin,
  Phys.\ Lett.\ B {\bf 94}, 245 (1980).

\bibitem{Bauer:2001yt} 
  C.~W.~Bauer, D.~Pirjol, and I.~W.~Stewart,
  Phys.\ Rev.\ D {\bf 65}, 054022 (2002)
  [hep-ph/0109045].

\bibitem{Dong:2011fb} 
  H.-R.~Dong, F.~Feng and Y.~Jia,
  JHEP {\bf 1110}, 141 (2011)
  [Erratum-ibid.\  {\bf 1302}, 089 (2013)]
  [arXiv:1107.4351 [hep-ph]].

\bibitem{Jia:2008ep}
  Y.~Jia and D.~Yang,
  Nucl.\ Phys.\ B {\bf 814} (2009) 217
  [arXiv:0812.1965 [hep-ph]].

\bibitem{Bauer:2000yr} 
  C.~W.~Bauer, S.~Fleming, D.~Pirjol, and I.~W.~Stewart,
  Phys.\ Rev.\ D {\bf 63}, 114020 (2001)
  [hep-ph/0011336].

\end{thebibliography}
\end{document}